\documentclass{AIAA}
\usepackage{nomencl}
\usepackage{subfigure}
\usepackage{amsmath} 
\usepackage{amssymb}
\usepackage[toc,page]{appendix}
\usepackage{color}
\usepackage{graphicx}

\usepackage{epstopdf}

\usepackage{graphics}
\makenomenclature

\begin{document}

\title{Continuum Deformation of a Multiple Quadcopter Payload Delivery Team without Inter-Agent Communication}

\author{Hossein Rastgoftar\footnote{Postdoctoral Research Fellow, Aerospace Engineering Department, University of Michigan Ann Arbor, Member AIAA.} and Ella M. Atkins\footnote{Professor, Aerospace Engineering Department, University of Michigan Ann Arbor, Associate Fellow AIAA.}}
\affiliation{University of Michigan, 1320 Beal Avenue, Ann Arbor, MI, USA 48109}

\begin{abstract}
This paper proposes continuum deformation as a strategy for controlling the collective motion of a multiple quadcopter system (MQS) carrying a common payload. Continuum deformation allows expansion and contraction of inter-agent distances in a $2-D$ motion plane to follow desired motions of three team leaders. The remaining quadcopter followers establish the desired continuum deformation only by knowing leaders' positions at desired sample time waypoints without the need for inter-agent communication over the intermediate intervals. 
Each quadcopter applies a linear-quadratic Gaussian (LQG) controller to track the desired trajectory given by the continuum deformation in the presence of disturbance and measurement noise. Results of simulated cooperative aerial payload transport in the presence of uncertainty illustrate application of continuum deformation for coordinated transport through a narrow channel.
\end{abstract}

\maketitle
\printnomenclature
\section{Introduction}
Cooperative control of multiple unmanned aerial vehicles (UAVs) has been an active area of research over the past two decades. Formation flight\cite{proud1999close, fax2004information}, air traffic control \cite{paterno1998formal}, transportation engineering \cite{michael2011cooperative}, aerobiological sampling over agricultural lands \cite{schmale2008development}, cooperative manipulation \cite{michael2009kinematics} and general team-based surveillance are some applications of cooperative control. A UAV team can improve mission efficiency, reduce cost, and offer  increased resilience to failures including individual UAV loss. Group cooperation also improves fault detection and ability for the team to recover from anomaly conditions \cite{bahceci2003review, bazoula2008formation, murray2007recent}. Consensus \cite{olfati2007consensus, ren2009distributed}, containment control \cite{cao2011distributed, li2012distributed} and partial differential equation (PDE) methods \cite{frihauf2011leader} offer schemes to  coordinate large-scale multi-UAV teams. 


 Cooperative payload transport, grasping and manipulation using multiple UAVs are applications that require UAVs to manage total forces and moments applied to external object(s) along with their collective motions. Manipulation can be used to achieve perching or deploy/pickup payloads, or the cooperative team can carry slung loads \cite{bernard2011autonomous, bernard2008slung}. Swing-free trajectory tracking is demonstrated by a single quadcopter carrying a payload in \cite{palunko2012trajectory}. Aerial manipulation using a single quadcopter is studied in \cite{kim2013aerial}. Modeling and control of multiple UAVs deployed for cooperative manipulation are investigated in \cite{michael2011cooperative, mellinger2011design, kim2013aerial, michael2009kinematics, sreenath2013dynamics}, while cooperative grasping of a payload using multiple UAVs is studied in \cite{mellinger2013cooperative, mellinger2011design}. Adaptive control for a slung load transport mission using a single rotorcraft is studied in \cite{bisgaard2010adaptive, potter2011reducing, dai2014adaptive}. Stability of slung load transport carrying by a multiple single-rotor helicopters is investigated in \cite{pounds2011grasping}, while an inverse kinematics formulation  for aerial payload transport is presented in \cite{jiang2013inverse}. 
  \begin{figure}
\center
\includegraphics[width=5.in]{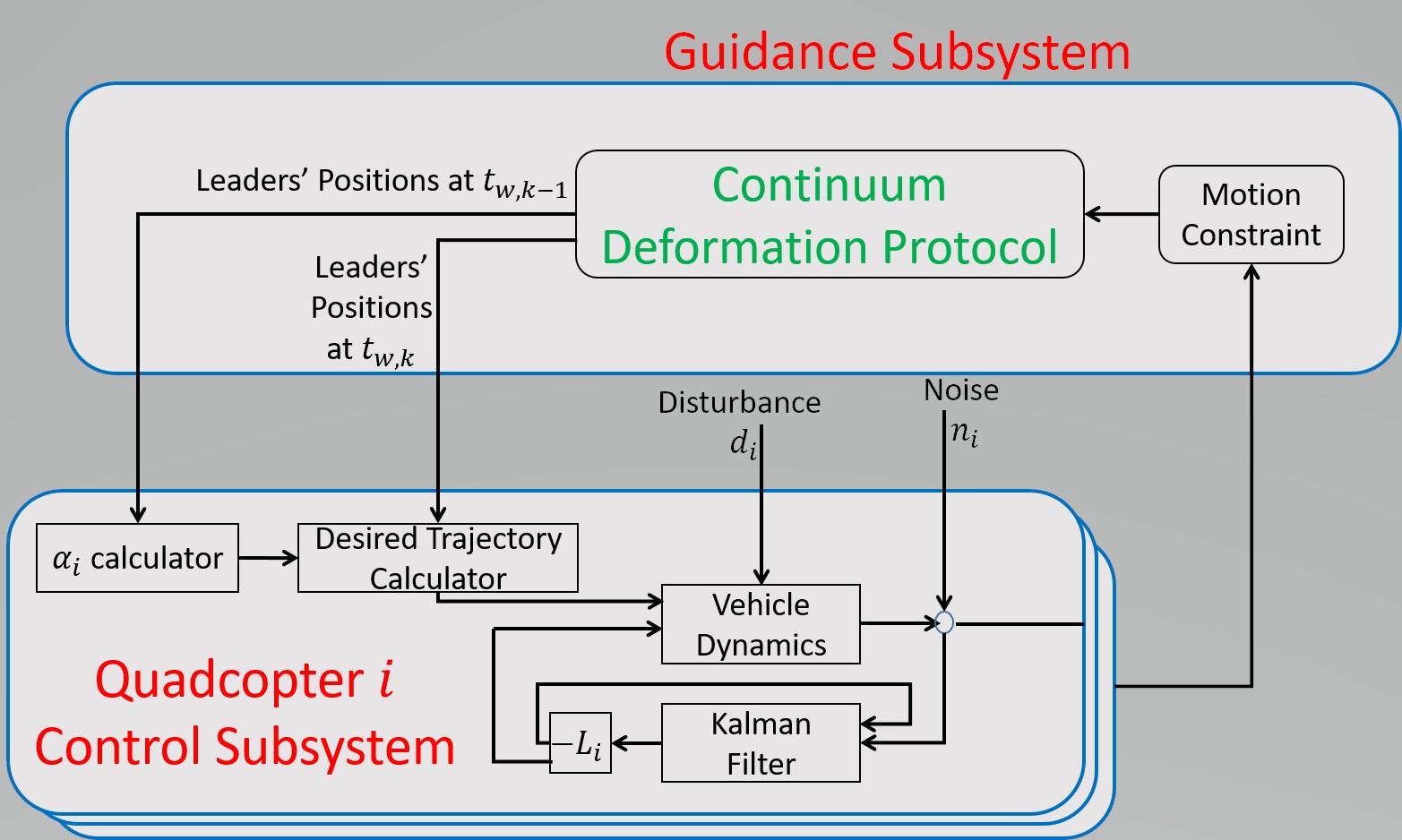}
\caption{Structure of the proposed method for aerial payload transport}
\label{Functionality}
\end{figure}
  
  This paper proposes a novel control strategy (Fig. \ref{Functionality}) for cooperative aerial payload transport in the presence of uncertainty.   A central guidance and control module manages leader UAV motions while follower UAV motions are efficiently regulated through continuum deformation.  A linear quadratic Gaussian (LQG) controller \cite{boyd1991linear} combining a linear quadratic controller and a Kalman estimator \cite{kalman1960new, brown1997introduction} offers robust trajectory tracking by each quadcopter. The guidance subsystem applies the recently proposed continuum deformation algorithm to achieve robust collective motion of quadcopter team during payload transport \cite{rastgoftar2016continuum, rastgoftarasymptotic}.  This protocol treats quadcopters as particles of a continuum and allows expansion and contraction of inter-agent distances. For continuum deformation in a plane, i.e., constant-altitude quadcopter motion, a leader-follower model designates three quadcopter leaders at the vertices of a triangle, called the \textit{leading triangle}.  The remaining quadcopters fly inside the leading triangle and are modeled as followers. 
  
  Application of continuum deformation to payload transport offers the following advantages:
  \begin{enumerate}
  \item{\textbf{Scalability}: The total number of quadcopters participating in a payload delivery mission can be chosen sufficiently high that each quadcopter in a multi-quadcopter system (MQS) can carry a modest payload weight.} 
  \item{\textbf{Minimal Inter-Agent Communication}:  Section \ref{Desired Trajectories of Quad-copters in a Continuum Deformation}  will describe how a desired MQS formation can be achieved without regular inter-agent communication. A follower quadcopter does not require position information of other follow quadcopters to learn the desired continuum deformation. Only leaders' positions at waypoint arrival times must communicated across the team, minimizing  communication costs and bandwidth requirements.}
  \item{\textbf{Negotiating obstacle-laden environments}: Because quadcopters are treated as particles of a continuum and inter-agent distances can be expanded or contracted via the continuum deformation strategy, the team can navigate constrained environments including passing through a narrow channel.  This capability will be demonstrated in simulation below.}
  \item{\textbf{Collision Avoidance}: Because the proposed continuum deformation protocol is guided by leaders at the vertices of a triangle and followers are all inside the leading triangle, collisions with obstacles can be avoided by choosing appropriate trajectories for leaders' motions. Due to the nonsingularity of continuum deformation, inter-agent collision can be avoided throughout the mission. }
  
  \end{enumerate}
  
  This article is organized as follows. First, multi-agent system continuum deformation is summarized in Section \ref{Preliminary Notions on a $2-D$ Continuum Deformation of Multi-Agent Systems}. Application to MQS continuum deformation for a cooperative payload delivery mission is described in Section \ref{Problem Statement}. The dynamics model of a single quadcopter as well as LQG controller design is described in Section \ref{Dynamics Model of a Single Quad-Copter}, while Section \ref{Desired Trajectories of Quad-copters in a Continuum Deformation} describes the guidance subsystem and formulates follower quadcopter desired trajectories learned without inter-agent communication. An aerial MQS payload transport case study in which the MQS negotiates a narrow channel is provided in Section \ref{Case Study: Cooperative Payload Delivery Negotiating a Narrow Channel} followed by concluding remarks in Section \ref{Conclusion}.
 





\section{$2-D$ Continuum Deformation of Multi-Agent Systems}
\label{Preliminary Notions on a $2-D$ Continuum Deformation of Multi-Agent Systems}
 \nomenclature{$V$}{A set defining index numbers of all quadcopters}\nomenclature{$V_F$}{A set defining index numbers of follower quadcopters}\nomenclature{$V_L$}{A set defining index numbers of leader quadcopters}


\label{Continuum Deformation of Multi-Agent Systems}
Let a continuum deformation in a $3-D$ motion space  be defined by 
\begin{equation}
t\geq t_0,~\mathbf{r}_{d,i}(t)=Q(t)\mathbf{r}_{d,i}(t_0)+\mathbf{D}(t)
\end{equation}
where $Q\in \mathbb{R}^{3\times 3}$ is the Jacobian matrix, $D\in \mathbb{R}^{3\times 1}$ is a rigid-body displacement vector, and
\begin{equation}
\label{HT}
\mathbf{r}_{d,i}(t)=x_{d,i}(t)\mathbf{I}+y_{d,i}(t)\mathbf{J}+z_{d,i}(t)\mathbf{K}.
\end{equation}
is called the \textit{desired position of agent $i$}  given by a homogeneous deformation. \nomenclature{$\mathbf{r}_{d,i}$}{Desired position of agent $i$}\nomenclature{$I_2\in \mathbb{R}^{2\times 2}$}{Identity matrix}\nomenclature{$\mathbf{1}_2\in \mathbb{R}^{2\times 1}$}{One-entry vector}
Without loss of generality, a $2-D$  continuum deformation in the $X-Y$ plane is considered in this article, therefore, 
\[
\forall t\geq t_0,~D_{3}(t)=0,~Q_{1,3}(t)=Q_{2,3}(t)=Q_{3,1}(t)=Q_{3,2}(t)=0,~Q_{3,3}(t)=1.
\]
For continuum deformation in a $2-D$ motion space, three leaders are placed at the vertices of a leading triangle and $V_L=\{1,2,3\}$ defines leader index numbers. Additionally, $N-3$ followers are located inside the leading triangle and $V_F=\{4\dots,N\}$ defines follower index numbers. Elements of the Jacobian matrix $Q$ and rigid-body displacement vector $\mathbf{D}$ can be uniquely related to leaders' positions by \cite{rastgoftar2016continuum, rastgoftarasymptotic, rastgoftar2017continuum}
\begin{equation}
\label{QD}
J_t
=
\begin{bmatrix}
I_2\otimes L_0&I_2\otimes \mathbf{1}_2
\end{bmatrix}
^{-1}
P_t
\end{equation}
 where "$\otimes$" is the Kronecker product symbol, $I_2\in \mathbb{R}^{2\times 2}$ is the identity matrix, and $\mathbf{1}_2\in \mathbb{R}^{2\times 1}$ is the one-entry vector, 
\begin{equation}
J_t=
\begin{bmatrix}
Q_{1,1}(t)&Q_{1,2}(t)&Q_{2,1}(t)&Q_{2,2}(t)&D_{1}(t)&D_{2}(t)
\end{bmatrix}
^T\in \mathbb{R}^{6\times 1},
\end{equation}
\begin{equation}
P_t=
\begin{bmatrix}
x_{d,1}(t)&x_{d,2}(t)&x_{d,3}(t)&y_{d,1}(t)&y_{d,2}(t)&y_{d,3}(t)
\end{bmatrix}
^T\in \mathbb{R}^{6\times 1},
\end{equation}
and
\begin{equation}
L_0=
\begin{bmatrix}
x_{d,1}(t_0)&y_{d,1}(t_0)\\
x_{d,2}(t_0)&y_{d,2}(t_0)\\
x_{d,3}(t_0)&y_{d,3}(t_0)\\
\end{bmatrix}
\in \mathbb{R}^{3\times 2}.
\end{equation}
\nomenclature{$x_{d,i}$}{$x$ component of the desired centroid position of quadcopter $i$}
\nomenclature{$y_{d,i}$}{$y$ component of the desired centroid position of quadcopter $i$}
\nomenclature{$z_{d,i}$}{$z$ component of the desired centroid position of quadcopter $i$}
Note that 
the matrix $[I_2\otimes L_0~I_2\otimes \mathbf{1}_2]$ is nonsingular if leader positions satisfy the following rank condition \cite{rastgoftar2016continuum, rastgoftarasymptotic, rastgoftar2017continuum}:
\begin{equation}
\forall~t\geq t_0,~
\begin{bmatrix}
\mathbf{r}_{d,2}-\mathbf{r}_{d,1}&\mathbf{r}_{d,3}-\mathbf{r}_{d,1}
\end{bmatrix}
=2.
\end{equation}
The continuum deformation defined in Eq. \eqref{HT} is called \textit{homogeneous deformation} because the Jacobian matrix $Q$ is only time-varying but is not spatially-varying.

\textbf{Invariant Parameters of a Homogeneous Deformation}: Under a homogeneous deformation,  the position of an agent $i$ can be expressed as the linear combination of leaders' positions,
\begin{equation}
\label{followersdesired}
\mathbf{r}_{d,i}(t)=\sum_{k=1}^3\alpha_{i,k}\mathbf{r}_{k}(t)
\end{equation}
where $\alpha_{i,k}$ is unique and obtained from
\begin{equation}
\label{alphaa}
\begin{bmatrix}
x_{d,1}(t_0)&x_{d,2}(t_0)&x_{d,3}(t_0)\\
y_{d,1}(t_0)&y_{d,2}(t_0)&y_{d,3}(t_0)\\
1&1&1\\
\end{bmatrix}
\begin{bmatrix}
\alpha_{i,1}\\
\alpha_{i,2}\\
\alpha_{i,3}\\
\end{bmatrix}
=
\begin{bmatrix}
x_{d,i}(t_0)\\
y_{d,i}(t_0)\\
1
\end{bmatrix}
.
\end{equation}
Parameters $\alpha_{i,1}$, $\alpha_{i,2}$, $\alpha_{i,3}$  remain unchanged at any time $t\geq t_0$, if $\mathbf{r}_{d,i}$ satisfies Eq. \eqref{HT} ($\forall i\in V$).\nomenclature{$\alpha_{i,k}$}{Invariant parameter of a homogeneous deformation ($i\in V_F,~k\in V_L$)}
Note that 
\begin{equation}
\sum_{k=1}^4\alpha_{i,k}=1
\end{equation}
and $\alpha_{i,k}>0$ when agent $i$ is inside the leading triangle.

\section{Problem Statement}
\label{Problem Statement}
Consider an MQS consisting of $N$ vehicles moving collectively  in a $3-D$ motion space. The MQS is deployed for payload aerial transport and it offers the ability of negotiating a narrow channel or other spatially-constrained environment. This paper proposes the architecture shown in Fig. \ref{Functionality} for guidance and control of the cooperative MQS. 

The guidance system applies principles of continuum mechanics \cite{lai2009introduction} to assign desired waypoints at certain sampling times given constraints of the mission. Note that desired waypoints provided by the guidance system are defined by a homogeneous deformation. The desired homogeneous deformation is formulated based on positions of three leader quadcopters at the vertices of the leading triangle as described above. The remaining quadcopters are followers acquiring the desired homogeneous deformation only by knowing leader waypoints at known sample times $t_0$, $t_{w,1}$, $\dots$, $t_{w,m}$, $t_f$. Section \ref{Desired Trajectories of Quad-copters in a Continuum Deformation} describes how the MQS desired formation given by a homogeneous deformation can be learned by followers without communication during intermediate time intervals $(t_0,t_{w,1})$, $(t_{w,1},t_{w,2})$, $\dots$, $(t_{w,m-1},t_{w,m})$, $(t_{w,m},t_{f})$.

The control system of an individual quadcopter uses a $6$ degree of freedom (DOF)  motion model. For control of each quadcopter, the  $12^{th}$ order quadcopter dynamics model is linearized around the desired state. Given the desired trajectory of quadcopter $i$ and the tension force exerted by a connecting cable,  the desired state of quadcopter $i$, $X_{d,i}$, is specified as described in Section \ref{AssigningDesiredState}.  An LQG controller offers robust tracking of the desired state $X_{d,i}$ given disturbances and measurement noise. The subsequent case study shows how an MQS consisting of $6$ quadcopters can cooperatively carrying a payload from an open area through a narrow channel.

\section{Quadcopter Dynamics Model}
\label{Dynamics Model of a Single Quad-Copter}
\subsection{Kinematics}
\subsubsection{Quadcopter Centroid Velocity and Acceleration}
The centroid position of quadcopter $i\in V$ is expressed with respect to an inertial frame attached to the ground (Earth):
\begin{equation}
\mathbf{r}_i=x_i\mathbf{I}+y_i\mathbf{J}+z_i\mathbf{K}
\end{equation}
\nomenclature{$x_i$}{$x$ component of the centroid position of quadcopter $i$}
\nomenclature{$y_i$}{$y$ component of the centroid position of quadcopter $i$}
\nomenclature{$z_i$}{$z$ component of the centroid position of quadcopter $i$}
Therefore, velocity and acceleration of quadcopter $i\in V$ is given by
\begin{equation}
\dot{\mathbf{r}}_i=\dot{x}_i\mathbf{I}+\dot{y}_i\mathbf{J}+\dot{z}_i\mathbf{K}
\end{equation}
\begin{equation}
\ddot{\mathbf{r}}_i=\ddot{x}_i\mathbf{I}+\ddot{y}_i\mathbf{J}+\ddot{z}_i\mathbf{K}
\end{equation}
\subsubsection{Quadcopter Centroid Angular Velocity and Acceleration}
Let $\phi_i$, $\theta_i$, $\psi_i$ be the "roll", "pitch", and "yaw" angles of quadcopter $i\in V$, then angular velocity of quadcopter $i$ is assigned by using the "3-2-1" standard 
Quadcopter angular velocity is given by
\begin{equation}
\label{AngOrig}
\mathbf{\Omega}_i=p_i\mathbf{i}_{b,i}+q_i\mathbf{j}_{b,i}+r_i\mathbf{k}_{b,i}=\dot{\psi}_i\mathbf{k}_{1,i}+\dot{\theta}_i\mathbf{j}_{2,i}+\dot{\phi}_i\mathbf{i}_{b,i},
\end{equation}
\nomenclature{$\phi_i$}{The roll angle of quadcopter $i$}
\nomenclature{$\theta_i$}{The pitch angle of quadcopter $i$}
\nomenclature{$\psi_i$}{The yaw angle of quadcopter $i$}
\nomenclature{$\mathbf{\Omega}_i$}{Angular velocity vector quadcopter $i$}
\nomenclature{$\mathbf{I},\mathbf{J},\mathbf{K}$}{Basis of the inertial coordinates affixed to the ground}
\nomenclature{$\mathbf{i}_{b,i},\mathbf{j}_{b,i},\mathbf{k}_{b,i}$}{Basis of the body coordinates for quadcopter $i$}
where
\[
\begin{split}
\begin{bmatrix}
\mathbf{i}_1\\
\mathbf{j}_1\\
\mathbf{k}_1\\
\end{bmatrix}
=&
\begin{bmatrix}
\cos\psi_i&\sin\psi_i&0\\
-\sin\psi_i&\cos\psi_i&0\\
0&0&1\\
\end{bmatrix}
\begin{bmatrix}
\mathbf{I}\\
\mathbf{J}\\
\mathbf{K}\\
\end{bmatrix}
,~
\begin{bmatrix}
\mathbf{i}_2\\
\mathbf{j}_2\\
\mathbf{k}_2\\
\end{bmatrix}
=
\begin{bmatrix}
\cos\theta_i&0&-\sin\theta_i\\
0&1&0\\
\sin\theta_i&0&\cos\theta_i\\
\end{bmatrix}
\begin{bmatrix}
\mathbf{i}_1\\
\mathbf{j}_1\\
\mathbf{k}_1\\
\end{bmatrix}
,\\
\begin{bmatrix}
\mathbf{i}_b\\
\mathbf{j}_b\\
\mathbf{k}_b\\
\end{bmatrix}
=&
\begin{bmatrix}
1&0&0\\
0&\cos\phi_i&\sin\phi_i\\
0&-\sin\phi_i&\cos\phi_i\\
\end{bmatrix}
\begin{bmatrix}
\mathbf{i}_2\\
\mathbf{j}_2\\
\mathbf{k}_2\\
\end{bmatrix}
.
\end{split}
\]
 Quadcopter angular velocity given in Eq. \eqref{AngOrig} can be rewritten in the following component-wise form:
\begin{equation}
\label{Wrelation}
\begin{bmatrix}
p_i\\
q_i\\
r_i
\end{bmatrix}
=W_{321,i}
\begin{bmatrix}
\dot{\phi}_i\\
\dot{\theta}_i\\
\dot{\psi}_i
\end{bmatrix}
\end{equation}
where
\begin{equation}
W_{321,i}=
\begin{bmatrix}
1&0&-\sin\theta_i\\
0&\cos\phi_i&\cos\theta_i\sin\phi_i\\
0&-\sin\phi_i&\cos\theta_i\cos\phi_i\\
\end{bmatrix}
.
\end{equation}
The angular acceleration of quadcopter $i$ is given by
\begin{equation}
\dot{\mathbf{\Omega}}_i=\dot{p}_i\mathbf{i}_{b,i}+\dot{q}_i\mathbf{j}_{b,i}+\dot{r}_i\mathbf{k}_{b,i}
\end{equation}

\subsection{Quadcopter Equations of Motion}
By applying Newton's second law, the quadcopter equations of motion become
\begin{equation}
\label{Newton2law}
\begin{split}
    \mathbf{F}_{Aero,i}+ \mathbf{F}_{Cord,i}-T_{i}\mathbf{k}_{b,i}-m_ig\mathbf{K}=m_i\ddot{\mathbf{r}}_i\\
    \mathbf{M}_i=I\dot{\mathbf{\Omega}}_i+\mathbf{\Omega}_i\times I_i\mathbf{\Omega}_i\\
\end{split}
,
\end{equation}
\nomenclature{$T_{i}$}{Thrust generated by quadcopter $i$}
\nomenclature{$m_i$}{Mass of quadcopter $i$}
where $m_i$ is the mass of quadcopter $i$, $g=9.81m/s^2$ is the gravitational acceleration, $\mathbf{F}_{Cord,i}$ is the tension force in the cord connecting quadcopter $i$ to the payload, and $T_i$ is the total thrust generated by quadcopter $i$'s rotors,
\[
I=
\begin{bmatrix}
I_{xx,i}&0&0\\
0&I_{yy,i}&0\\
0&0&I_{zz,i}
\end{bmatrix}
,~
\mathbf{F}_{Aero,i}=
\begin{bmatrix}
-A_{x,i}\dot{x}\\
-A_{y,i}\dot{y}\\
-A_{z,i}\dot{x}\\
\end{bmatrix}
,~\mathbf{M}_i=
\begin{bmatrix}
\tau_{\phi,i}\\
\tau_{\theta,i}\\
\tau_{\psi,i}\\
\end{bmatrix}
,~
\mathbf{F}_{Cord,i}=
\begin{bmatrix}
P_{x,i}\\
P_{y,i}\\
P_{z,i}\\
\end{bmatrix}
,
\]
$A_{x,i}$, $A_{y,i}$, $A_{z,i}$ represent aerodynamics parameters. Expressing $\mathbf{k}_{b,i}$ with respect to $\mathbf{I}$,  $\mathbf{J}$, and  $\mathbf{K}$,
\begin{equation}
\label{KKBBII}
\begin{split}
\mathbf{k}_{b,i}=&-\sin\phi_i\mathbf{j}_{2,i}+\cos\phi_i\mathbf{k}_{2,i}\\
=&-\sin\phi_i\mathbf{j}_{1,i}+\cos\phi_i\left(\sin\theta_i\mathbf{i}_{1,i}+\cos\theta_i\mathbf{k}_{1,i}\right)\\
=&-\sin\phi_i\left(-\sin\psi_i\mathbf{I}+\cos\psi_i\mathbf{J}\right)+\cos\phi_i\sin\theta_i\left(\cos\psi_i\mathbf{I}+\sin\psi_i\mathbf{J}\right)+\cos\phi_i\sin\theta_i\mathbf{K}\\
=&\mathbf{I}\left(\cos\phi_i\sin\theta_i\cos\psi_i+\sin\phi_i\sin\psi_i\right)+
\mathbf{J}\left(\cos\phi_i\sin\theta_i\sin\psi_i-\sin\phi_i\cos\psi_i\right)+
\mathbf{K}\cos\theta_i\cos\phi_i
\end{split}
,
\end{equation}
equations of motion of quadcopter $i$ can be rewritten in the following form:
\begin{equation}
\label{quadcopterdyn}
\dot{X}_i=f_i(X_i)+g_i(X_i)U_i
\end{equation}
where
\[
X_i=
\begin{bmatrix}
{x}_i&{y}_i&{z}_i&{\phi}_i&{\theta}_i&{\psi}_i&{u}_i&{v}_i&{w}_i&{p}_i&{q}_i&{r}_i
\end{bmatrix}
^T
\in \mathbb{R}^{12\times 1}
,
\]
\[
U_i=
\begin{bmatrix}
T_i&
\tau_{\phi,i}&
\tau_{\theta,i}&
\tau_{\psi,i}
\end{bmatrix}
^T\in \mathbb{R}^{4\times 1}
,
\]
\[
\begin{split}
f_i=
\begin{bmatrix}
u_i\\
v_i\\
w_i\\
p_i+q_i\dfrac{\sin\phi_i\sin\theta_i}{\cos\theta_i}+r\dfrac{\cos\phi_i\sin\theta_i}{\cos\theta_i}\\
q_i\cos\phi_i-r_i\sin\phi_i\\
q_i\dfrac{\sin\phi_i}{\cos\theta_i}+r_i\dfrac{\cos\phi_i}{\cos\theta_i}\\
P_{x,i}-\dfrac{A_{x,i}}{m_i}u_i\\
P_{y,i}-\dfrac{A_{y,i}}{m_i}v_i\\
P_{z,i}-\dfrac{A_{z,i}}{m_i}w_i\\
\dfrac{I_{yy,i}-I_{zz,i}}{I_{xx,i}}q_ir_i\\
\dfrac{I_{zz,i}-I_{xx,i}}{I_{yy,i}}r_ip_i\\
\dfrac{I_{xx,i}-I_{yy,i}}{I_{zz,i}}p_iq_i\\
\end{bmatrix}
,~g_iU_i=
\begin{bmatrix}
0\\
0\\
0\\
0\\
0\\
0\\
\dfrac{T_i}{m_i}\left(\cos\psi_i\sin\theta_i\cos\phi_i+\sin\psi_i\sin\phi_i\right)\\
\dfrac{T_i}{m_i}\left(\sin\psi_i\sin\theta_i\cos\phi_i-\cos\psi_i\sin\phi_i\right)\\
\dfrac{T_i}{m_i}\left(\cos\theta_i\cos\phi_i\right)\\
\dfrac{\tau_{\phi,i}}{I_{xx,i}}\\
\dfrac{\tau_{\theta,i}}{I_{yy,i}}\\
\dfrac{\tau_{\psi,i}}{I_{zz,i}}\\
\end{bmatrix}
.
\end{split}
\]
\subsection{Quadcopter LQG Control Law}
An LQG controller offers robust tracking of the desired state $X_{d,i}$ for each quadcopter $i$ in the presence of uncertainty.  Section \ref{AssigningDesiredState} formulates the desired state and input given a desired trajectory $\mathbf{r}_{d,i}$ assigned by continuum deformation. An LQG controller is then designed in Section \ref{LQG Controller}  to track the desired position defined by the continuum deformation in the presence of uncertainty.
\subsubsection{Assigning Desired State $X_{d,i}$ and Desired Input  $U_{d,i}$}
\label{AssigningDesiredState}
The $x_{d,i}$, $y_{d,i}$, $z_{d,i}$, $u_{d,i}$, $v_{d,i}$, and $w_{d,i}$ of desired state,
\[
X_{d,i}=
\begin{bmatrix}
{x}_{d,i}&{y}_{d,i}&{z}_{d,i}&{\phi}_{d,i}&{\theta}_{d,i}&{\psi}_{d,i}&{u}_{d,i}&{v}_{d,i}&{w}_{d,i}&{p}_{d,i}&{q}_{d,i}&{r}_{d,i}
\end{bmatrix}
^T
\in \mathbb{R}^{12\times 1}
,
\]
are obtained from the guidance system by applying the proposed continuum deformation protocol as described below in Section \ref{Desired Trajectories of Quad-copters in a Continuum Deformation}.

The following procedure is used ito  assign $X_{d,i}$ and $U_{d,i}=[T_{d,i}~\tau_{\phi,d,i}~\tau_{\theta,d,i}~\tau_{\psi,d,i}]^T\in \mathbb{R}^{4\times 1}$:
\begin{enumerate}
\item{\textbf{Assigning $\psi_{d,i}$}: Given the desired trajectory
$\mathbf{r}_{d,i}=x_{d,i}(t)\mathbf{I}+y_{d,i}(t)\mathbf{J}+z_{d,i}(t)\mathbf{K}$, $\dot{\mathbf{r}}_{d,i}=\dot{x}_{d,i}(t)\mathbf{I}+\dot{y}_{d,i}(t)\mathbf{J}+\dot{z}_{d,i}(t)\mathbf{K}$, and $\ddot{\mathbf{r}}_{d,i}=\ddot{x}_{d,i}(t)\mathbf{I}+\ddot{y}_{d,i}(t)\mathbf{J}+z_{d,i}(t)\mathbf{K}$ the desired yaw angle $\psi_{d,i}(t)$ is obtained.}
\item{\textbf{Assigning $T_{d,i}$, $\phi_{d,i}$ and $\theta_{d,i}$}: 
Substituting $\dot{x}_{i}(t)$, $\dot{y}_{i}(t)$, $\dot{z}_{i}(t)$, $\dot{\psi}_{i}(t)$ in Eq. \eqref{KKBBII} by $\dot{x}_{d,i}(t)$, $\dot{y}_{d,i}(t)$, $\dot{z}_{d,i}(t)$, $\dot{\psi}_{d,i}(t)$, $T_{d,i}$ (desired tension in the connecting rope) and $\mathbf{k}_{b,d,i}(t)$ (desired orientation of $\mathbf{k}_{b,i}(t)$), are obtained as follows:
\begin{equation}
\label{THRUST}
T_{d,i}=||\mathbf{F}_{Aero,d,i}+\mathbf{F}_{Cord,i}-m_ig\mathbf{K}-m_i\ddot{\mathbf{r}}_{d,i}||
\end{equation}
\begin{equation}
\label{kbi}
\mathbf{k}_{b,d,i}=\dfrac{\mathbf{F}_{Aero,d,i}+\mathbf{F}_{Cord,i}-m_ig\mathbf{K}-m_i\ddot{\mathbf{r}}_{d,i}}{T_{d,i}}.
\end{equation}

From  $\mathbf{k}_{b,d,i}$ obtained per Eq. \eqref{kbi}, 
\[
\mathbf{k}_{b,d,i}=
\begin{bmatrix}
b_{x,i}\\
b_{y,i}\\
b_{z,i}\\
\end{bmatrix}
,
\]
$\phi_{d,i}(t)$ and $\theta_{d,i}(t)$ are obtained as follows:
\begin{equation}
\label{fidthetad}
\begin{split}
\phi_{d,i}=\arcsin \left(b_{x,i}\cos\psi_{d,i}-b_{y,i}\sin\psi_{d,i}\right)\\
\theta_{d,i}=\tan ^{-1}\left(\dfrac{b_{x,i}\sin\psi_{d,i}+b_{y,i}\sin\psi_{d,i}}{b_{z,i}}\right)\\
\end{split}
.
\end{equation}
}
\item{\textbf{Assigning $p_{d,i}$, $q_{d,i}$, and $r_{d,i}$}: Given $\phi_{d,i}$, $\theta_{d,i}$, and $\psi_{d,i}$ at discrete times $t_k$ and $t_{k-1}$, $\dot{\phi}_{d,i}$, $\dot{\theta}_{d,i}$, and $\dot{\psi}_{d,i}$ are numerically assigned:
\begin{equation}
\label{fidthetadsaid}
\begin{split}
\dot{\phi}_{d,i}(t_k)=\dfrac{\phi_{d,i}(t_k)-\phi_{d,i}(t_{k-1})}{t_k-t_{k-1}}\\
\dot{\theta}_{d,i}(t_k)=\dfrac{\theta_{d,i}(t_k)-\theta_{d,i}(t_{k-1})}{t_k-t_{k-1}}\\
\dot{\psi}_{d,i}(t_k)=\dfrac{\psi_{d,i}(t_k)-\psi_{d,i}(t_{k-1})}{t_k-t_{k-1}}\\
\end{split}
\end{equation}
$\dot{\phi}_{d,i}$, $\dot{\theta}_{d,i}$, and $\dot{\psi}_{d,i}$ are then related to $p_{d,i}$, $q_{d,i}$, and $r_{d,i}$ by applying the relation \eqref{Wrelation}, where $\phi_i$, $\theta_i$, and $\psi_i$ in $W_{321}$ are replaced by $\phi_{d,i}$, $\theta_{d,i}$, and $\psi_{d,i}$, respectively.}
\item{\textbf{Assigning $\tau_{\phi,d,i}$, $\tau_{\theta,d,i}$ and $\tau_{\psi,d,i}$}: By knowing $p_{d,i}$, $q_{d,i}$, and $r_{d,i}$ at discrete times $t_{k-1}$ and $t_k$, $\dot{p}_{d,i}$, $\dot{q}_{d,i}$, and $\dot{r}_{d,i}$ are numerically obtained from
\begin{equation}
\label{fidthetadsaid}
\begin{split}
\dot{p}_{d,i}(t_k)=\dfrac{p_{d,i}(t_k)-p_{d,i}(t_{k-1})}{t_k-t_{k-1}}\\
\dot{q}_{d,i}(t_k)=\dfrac{q_{d,i}(t_k)-q_{d,i}(t_{k-1})}{t_k-t_{k-1}}\\
\dot{r}_{d,i}(t_k)=\dfrac{r_{d,i}(t_k)-r_{d,i}(t_{k-1})}{t_k-t_{k-1}}\\
\end{split}
.
\end{equation}
Then, $\tau_{\phi,d,i}$, $\tau_{\theta,d,i}$, and  $\tau_{\psi,d,i}$ are assigned from Eq. \eqref{Newton2law}.
}
\end{enumerate}

\label{AssigningDesiredInput}


\subsubsection{LQG Controller}
\label{LQG Controller}
To find control $U_i$ at $t\in [t_{k-1},t_k]$, we use a linearized model of the quadcopter dynamics,
\begin{equation}
\begin{split}
\dot{\delta X}_{i,k}=&A_{i,k}\delta X_{i,k}+B_{i,k}\delta U_{i,k}+d_{i,k}\\
\delta Y_{i,k}=&C_{i,k}\delta X_{i,k}+n_{i,k}\\
\end{split}
\end{equation}
where $C_{i,k}\in \mathbb{R}^{12\times 12}$ is the identity matrix, $d_{i,k}\in \mathbb{R}^{12\times 1}$ is a zero-mean distarbance, $n_{i,k}\in \mathbb{R}^{12\times 1}$ is a zero-mean noise, and 
\begin{equation}
\label{linearization}
\begin{split}
A_{i,k}=\dfrac{\partial}{\partial X_{d,i}(t_k)}\bigg[ f_i(X_{d,i}(t_k))+g_i(X_{d,i}(t_k))U_{d,i}(t_k)\bigg]\\
B_{i,k}=\dfrac{\partial}{\partial U_{d,i}(t_k)}\bigg[ f_i(X_{d,i}(t_k))+g_i(X_{d,i}(t_k))U_{d,i}(t_k)\bigg]\\
\end{split}
.
\end{equation}
\nomenclature{$d_{i,k}\in \mathbb{R}^{12\times 1}$}{Disturbance}
\nomenclature{$n_{i,k}$}{Measurement noise}
\nomenclature{$L_{i,k}$}{Kalman gain}
\nomenclature{$K_{i,k}$}{LQ control gain}
Note that 
\begin{equation}
\label{Discrete}
\begin{split}
X_i(t)=X_{d,i}(t_{k-1})+\delta X_{i,k}(t-t_{k-1})\\
U_i(t)=U_i(t_{k-1})+\delta U_{i,k}(t-t_{k-1})\\
\end{split}
\end{equation}
are the actual control state and input of the quadcopter dynamics \eqref{quadcopterdyn}. Given $X_i(t_{k-1})$,
\begin{equation}
\label{initialtimevarying}
\delta{X}_{k,i,0}=X_i(t_{k-1})-X_{d,i}(t_{k-1})
\end{equation}
is considered as the initial state of the quadcopter $i$ at the time $t_{k-1}$

  \begin{figure}
\center
\includegraphics[width=5.in]{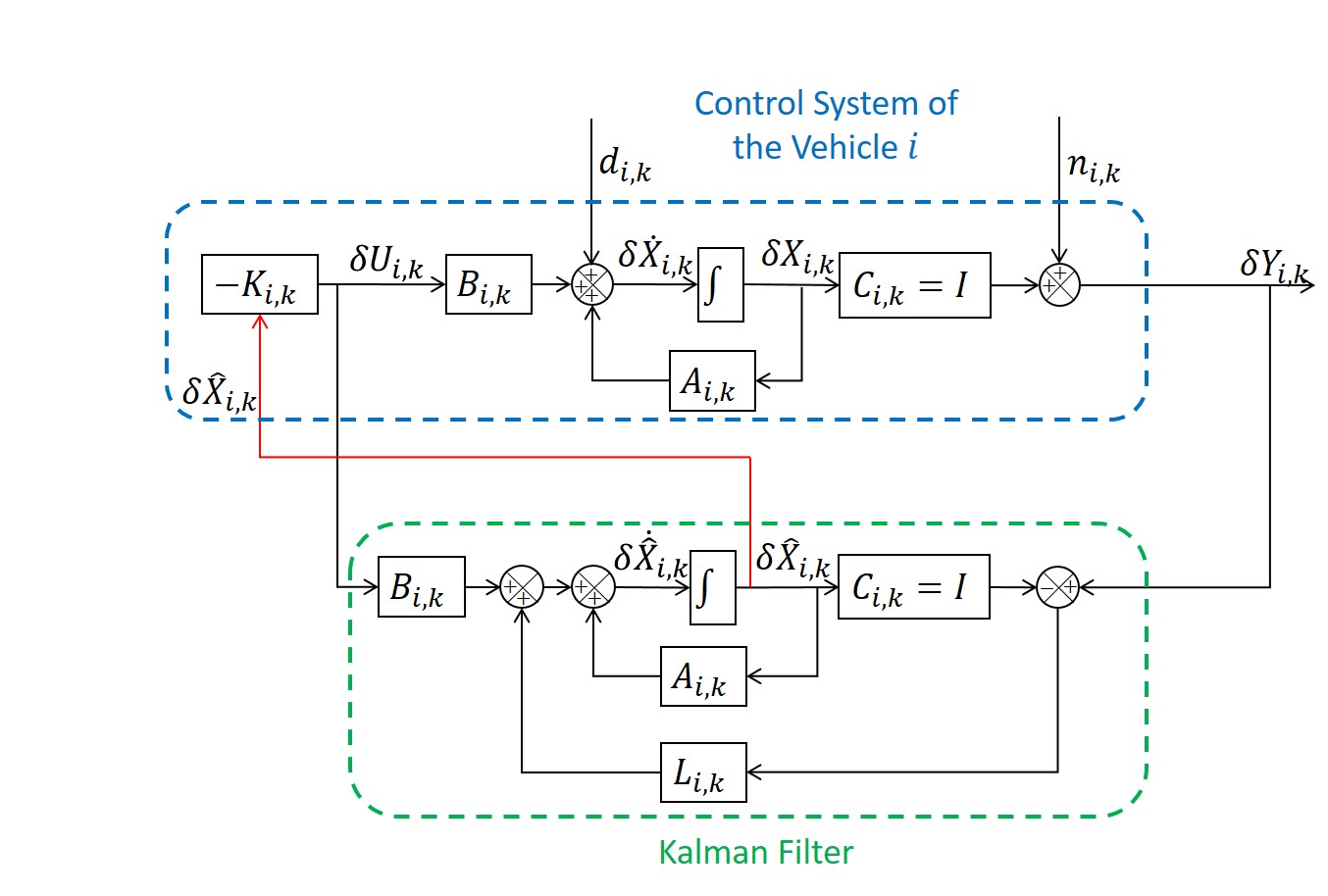}
\caption{Structure of the LQG controller of the quadcopter $i$ integrating a linear quadratic controller and a Kalman estimator}
\label{KALMAN}
\end{figure}
  

\textbf{LQG Controller}: As shown in Fig. \ref{KALMAN},  LQG design integrates a linear quadratic (LQ) controller and a Kalman estimator to the control system of the the quadcopter $i\in V_F$.

The Kalman filter dynamics
\begin{equation}
\delta \dot{\hat{X}}_{i}=A_{i,k}\delta {\hat{X}}_{i}+B_{i,k}\delta {{U}}_{i,k}+L_{i,k}(\delta Y_{i,k})
\end{equation}
where 
\begin{equation}
L_{i,k}=P_{i,k}{C_{i,k}}^T{R_{i,k}}^{-1}
\end{equation}
is called a \textit{Kalman gain} and obtained from the Riccati equation \cite{bittanti2012riccati},
\begin{equation}
\begin{split}
0=&A_{i,k}P_{i,k}+P_{i,k}{A_{i,k}}^T-P_{i,k}{C_{i,k}}^T{R_{i,k}}^{-1}C_{i,k}P_{i,k}+Q_{i,k}\\
Q_{i,k}=&\mathbb{E}(d_{i,k}{d_{i,k}}^T)\\
R_{i,k}=&\mathbb{E}(n_{i,k}{n_{i,k}}^T)\\
\end{split}
\end{equation}

Also, the control is chosen as
\begin{equation}
U_{i,k}=K_{i,k}\delta \hat{X}_{i,k}
\end{equation}
where 
\begin{equation}
K_{i,k}=-{H_{i,k}}^{-1}{B_{i,k}}^TS_{i,k}
\end{equation}
is obtained by solving algebraic Riccati equation:
\begin{equation}
\begin{split}
0=&S_{i,k}{A_{i,k}}+{A_{i,k}}^T{S_{i,k}}-S_{i,k}{B_{i,k}}{H_{i,k}}^{-1}{B_{i,k}}^TS_{i,k}+E_{i,k}\\
H_{i,k}\in&\mathbb{R}^{4\times 4}>0\\
E_{i,k}\in&\mathbb{R}^{12\times 12}\geq 0\\
\end{split}
.
\end{equation}
Note that $H_{i,k}=\mathbb{E}(n_{i,k}{n_{i,k}}^T)$ and $E_{i,k}=\mathbb{E}(d_{i,k}{d_{i,k}}^T)$ where an  $\mathbb{E}(\cdot)$ denotes the expected value. 

\section{Desired Trajectories of Quadcopters in a Continuum Deformation}
\label{Desired Trajectories of Quad-copters in a Continuum Deformation}
Suppose every follower quadcopter $i$ ($\forall i\in V_F$) knows its own position and leaders' positions at known sampling time $t_{w,k-1}$ and $t_{w,k}$. Then, every follower quadcopter $i$ can set up its own desired trajectory without communication with other followers by applying the following steps:
\begin{enumerate}
\item{\textbf{$\alpha_i$ Calculation}: Assign parameters $\alpha_{i,1}(t_{w,k-1})$, $\alpha_{i,2}(t_{w,k-1})$, and $\alpha_{i,3}(t_{w,k-1})$ by using the Eq. \eqref{alphaa} (In Eq. \eqref{alphaa} $t_0$ should be replaced by $t_{w,k-1}$.)}
\item{\textbf{Desired Trajectory Calculation}: Set the desired trajectory of the follower $i\in V_F$ at $t\in [t_{w,k-1},t_{w,k-1}]$ given leaders' initial and final positions:
\begin{equation}
\label{desiredtrajectory}
i\in V_F,~\mathbf{r}_{d,i}(t)=\sum_{k=1}^3\alpha_{i,k}\bigg[\dfrac{t-t_{w,k-1}}{t_{w,k}-t_{w,k-1}}\left(\mathbf{r}_k(t_{w,k})-\mathbf{r}_k(t_{w,k-1})\right)+\mathbf{r}_k(t_{w,k-1})\bigg]
\end{equation}
.}
\end{enumerate}

Note that the desired velocity of the follower $i\in V_F$ is constant:
\begin{equation}
\label{desiredtrajectory}
i\in V_F,
\begin{bmatrix}
 u_{d,i}\\
v_{d,i}\\
v_{d,i}\\
\end{bmatrix}
=
\begin{bmatrix}
 \sum_{k=1}^3\alpha_{i,k}\bigg\{\dfrac{\left(x_k(t_{w,k})-x_k(t_{w,k-1})\right)}{t_{w,k}-t_{w,k-1}}\bigg\}\\
\sum_{k=1}^3\alpha_{i,k}\bigg\{\dfrac{\left(y_k(t_{w,k})-y_k(t_{w,k-1})\right)}{t_{w,k}-t_{w,k-1}}\bigg\}\\
\sum_{k=1}^3\alpha_{i,k}\bigg\{\dfrac{\left(z_k(t_{w,k})-z_k(t_{w,k-1})\right)}{t_{w,k}-t_{w,k-1}}\bigg\}\\
\end{bmatrix}
.
\end{equation}

\section{Case Study: Cooperative Payload Delivery through a Narrow Channel}
\label{Case Study: Cooperative Payload Delivery Negotiating a Narrow Channel}

This section considers an MQS consisting of $20$ quadcopters, three leaders and seventeen followers. The MQS negotiates a narrow channel while it carries a payload. Leader and follower quadcopters all have the same properties as listed in Table \ref{tb:const} \cite{luukkonen2011modelling}. 

\subsection{Desired Continuum Deformation}
Leaders are located at $(x_{d,1}(t_0),y_{d,1}(t_0),z_{d,1}(t_0))=(-20,-20,50)$, $(x_{d,2}(t_0),y_{d,2}(t_0),z_{d,2}(t_0))=(20,-18,50)$, and $(x_{d,3}(t_0),y_{d,3}(t_0),z_{d,3}(t_0))=(0,-20,50)$ at $t_0=0s$. Leaders choose $(x_{d,1}(t_f),y_{d,1}(t_f),z_{d,1}(t_f))=(-15,0,50)$, $(x_{d,2}(t_0),y_{d,2}(t_0),z_{d,2}(t_0))=(15,10,50)$, and $(x_{d,3}(t_0),y_{d,3}(t_0),z_{d,3}(t_0))=(0,35,50)$ as their desired positions at $t_f=20s$. 
The desired trajectories of the leaders are defined by
\begin{equation}
Leader~1~,\mathbf{r}_{1,d}(t)=
\begin{bmatrix}
-5\\
20\\
0\\
\end{bmatrix}
\dfrac{t}{20}+
\begin{bmatrix}
20\\
-20\\
50\\
\end{bmatrix}
\end{equation}
\begin{equation}
Leader~2~,\mathbf{r}_{2,d}(t)=
\begin{bmatrix}
0\\
15\\
0
\end{bmatrix}
\dfrac{t}{20}+
\begin{bmatrix}
0\\
20\\
50\\
\end{bmatrix}
\end{equation}
\begin{equation}
Leader~3~,\mathbf{r}_{3,d}(t)=
\begin{bmatrix}
-5\\
28\\
0
\end{bmatrix}
\dfrac{t}{20}+
\begin{bmatrix}
20\\
-18\\
50
\end{bmatrix}
\end{equation}
\begin{table}[h]
     \centering
     \caption{Quadcopter Simulation Parameters \cite{luukkonen2011modelling}}
     \begin{tabular}{c c}
     \hline
         Parameter & Value\\
         \hline
         \hline
         $g$&$9.81~m/s^2$ \\
         $m$&$0.468~kg$\\
         $I_{xx}$&$4.856\times 10^{-3}~kg\cdot m^2$ \\
         $I_{yy}$&$4.856\times 10^{-3}~kg\cdot m^2 $\\
         $I_{zz}$&$8.801\times 10^{-3}~kg\cdot m^2$\\
         $A_x$&$0.25~kg/s$\\
         $A_y$&$0.25~kg/s$\\
         $A_z$&$0.25~kg/s$\\
         \hline
     \end{tabular}
     \label{tb:const}
 \end{table}

\begin{figure}
\centering
\subfigure[]{\includegraphics[width=0.45\linewidth]{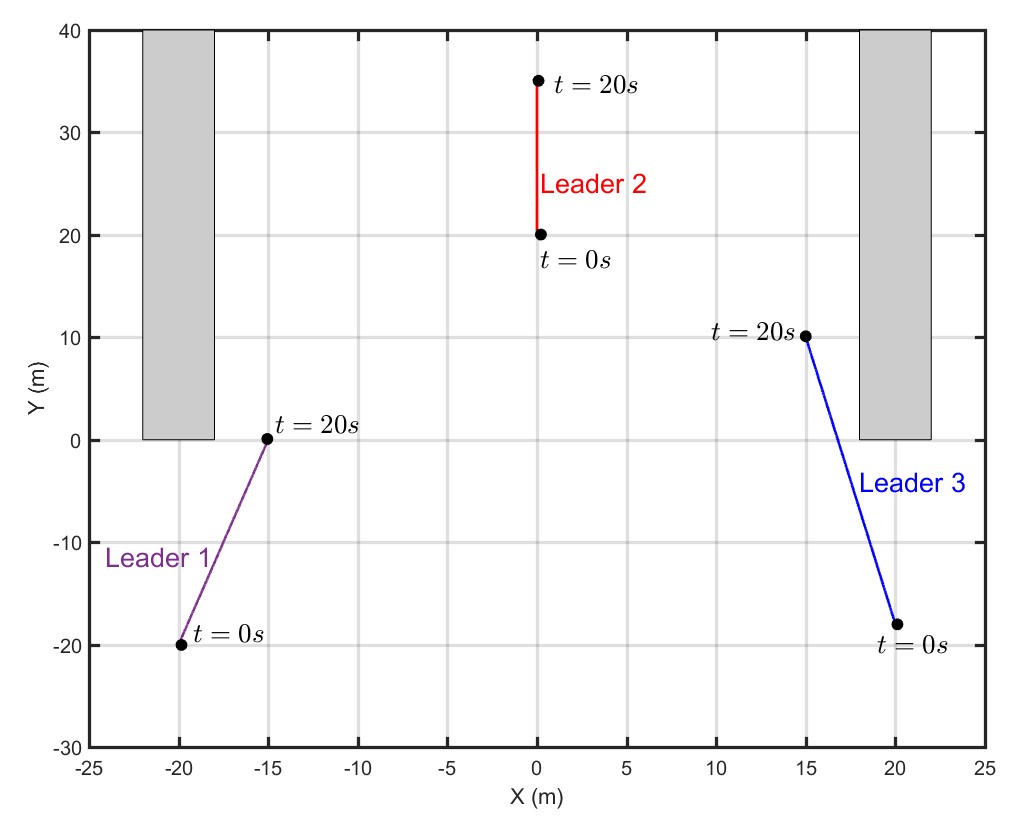}}
\subfigure[]{\includegraphics[width=0.45\linewidth]{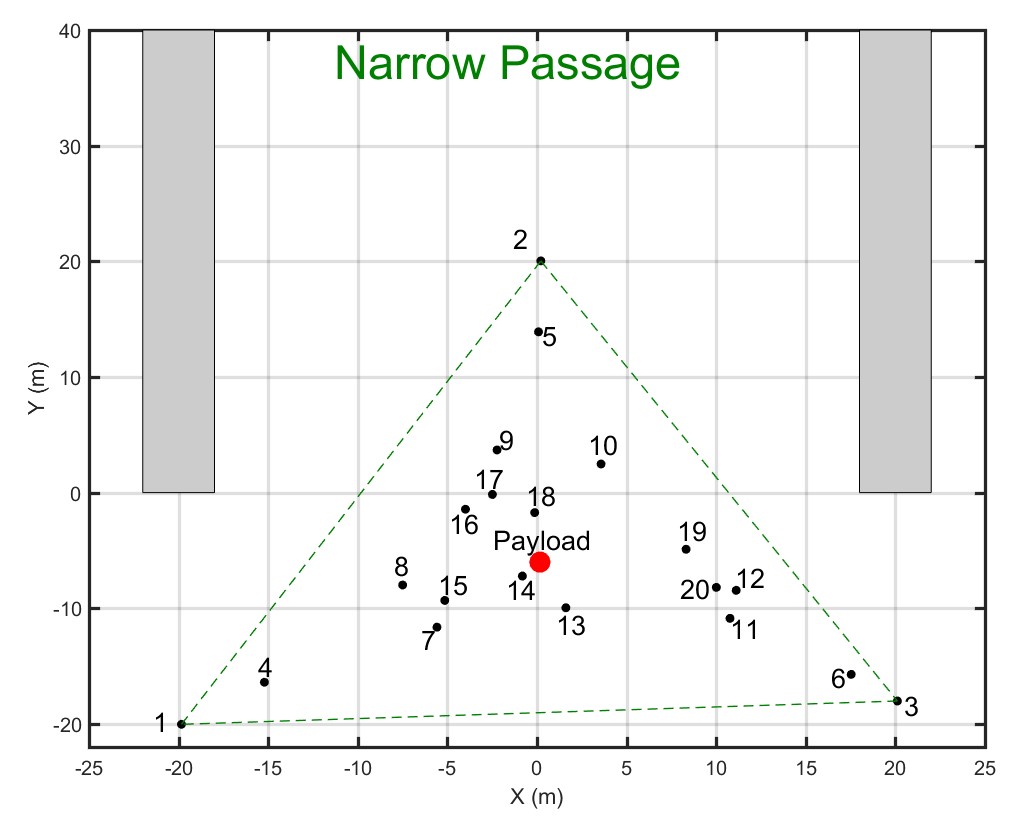}}
\caption{(a) Top view of leader quadcopter paths in the plane $Z=50$; (b) Inter-agent communication among quadcopters with quadcopters depicted in their initial positions.}
\label{Leaderspaths}
\end{figure}

 Fig. \ref{Leaderspaths} (a) shows a top view of leaders' paths in the plane $Z=50m$. Given leaders' positions at $t\in [0,20]$, elements of the Jacobian $Q$ and rigid-body translation vector $\mathbf{D}$ of the desired continuum deformation are obtained from Eq. \eqref{QD} as depicted in Fig. \ref{schematicpayloaddelivery}  versus time. As shown, $Q_{1,3}(t)=Q_{3,1}(t)=Q_{2,3}(t)=Q_{3,2}(t)=0$ and $Q_{3,3}(t)=1$ ($\forall t\in [0,20]$) imply continuum deformation of the MQS in a plane normal to the $Z$ axis. Additionally, $Q(0)=I_3$ and $\mathbf{D}(0)=\mathbf{0}\in \mathbb{R}^{3\times 1}$.
 
 \begin{figure}
\centering
\subfigure[Elements of $Q$]{\includegraphics[width=0.45\linewidth]{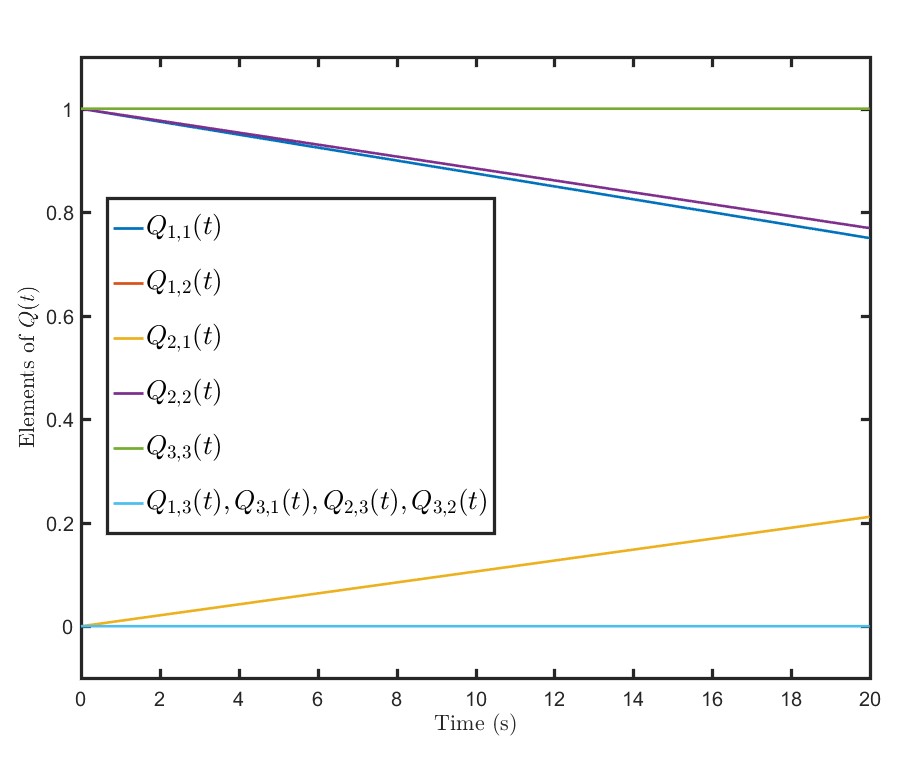}}
\subfigure[Elements of $D$]{\includegraphics[width=0.45\linewidth]{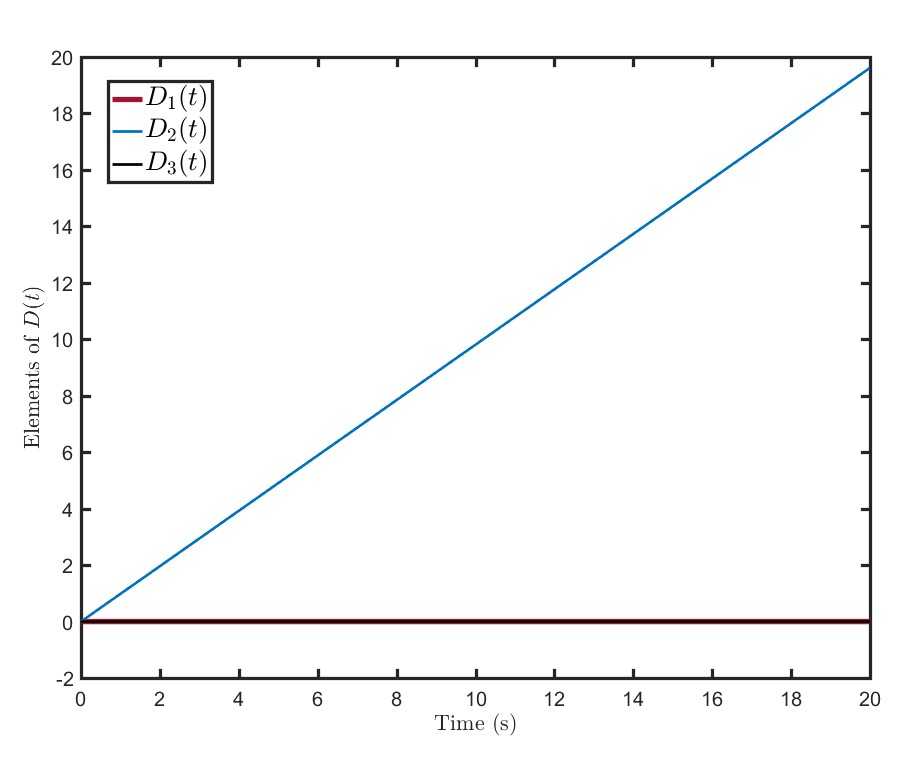}}
\caption{Elements of $Q$ and $D$ versus time}
\label{schematicpayloaddelivery}
\end{figure}
 
 Each quadcopter learns the desired continuum deformation without communicating during $t\in [0,20]$. Each follower $i\in V_F$ only needs to know the leaders' initial and final positions as well as its own initial positions. Given initial position of follower $i$ as well as leaders' initial positions (See Fig.  \ref{schematicpayloaddelivery} (b).), parameter $\alpha_{i,k}$ ($i\in V_F=\{4,5,6\},~k\in V_L=\{1,2,3\}$) is assigned from Eq. \eqref{alphaa}. Parameters $\alpha_{i,k}$ as well as quadcopter initial positions are listed in Table \ref{Table2}. Knowing parameters $\alpha_{i,1}$, $\alpha_{i,2}$, $\alpha_{i,3}$, as well as leaders' initial and final desired positions, follower $i$ can assign its own desired trajectory according to Eq. \eqref{desiredtrajectory}.
 
 \begin{table}[h]
\centering
\caption{Quadcopter positions at $t=t_0$; parameters $\alpha_{i,k}$ ($i\in V_F,~k\in V_L$)}
\label{Table2}
\begin{center}
\begin{tabular}{c c c c c c c }
\hline
i&$x_i(t_0)$&$y_i(t_0)$&$z_i(t_0)$&$\alpha_{i,1}$&$\alpha_{i,2}$&$\alpha_{i,3}$\\
\hline
\hline
1& -20   & -20     &50    &-&-&-\\
2& 0     & 20      &50   &-&-&-\\
3& 20    & -18     &50    &-&-&-\\
4& 18.5553&   -16.4474&50  &    0.8411&     0.0838 &    0.0751\\
5&    2.9446  &  14.0859&50 &     0.0774   &  0.8497  &    0.0730\\
6&   18.7505&   -15.5800 &50  &   0.0334 &    0.0625  &    0.9040\\
7&    15.8793&   -11.5596  &50  &  0.5457  &   0.1989 &    0.2554\\
8&    14.2071  &  -7.9219&50 &     0.5462&     0.2937 &    0.1600\\
9&     8.1559&     3.7254&50 &     0.2730  &   0.5857   &  0.1413\\
10&    9.0793   &  2.5421&50 &     0.1431&     0.5475   &  0.3094\\
11&    16.1245&   -10.9749&50 &     0.1365   &  0.1936    & 0.6699\\
12&    14.7419 &   -8.5407&50 &     0.0996 &    0.2578&     0.6426\\
13&   15.3257&    -9.5906 &50 &    0.3499&     0.2343   &  0.4158\\
14&   13.8798  &  -7.1498 &50 &    0.3720    & 0.3052&     0.3228\\
15&   14.9875&    -9.0965 &50 &    0.5025  &   0.2577  &   0.2398\\
16&   10.9917  &  -1.7927 &50 &    0.3838&     0.4547    & 0.1615\\
17&   10.4800  &  -0.2296 &50 &    0.3252  &   0.4864&     0.1884\\
18&   10.9509&    -1.7695 &50 &    0.2883&     0.4425 &    0.2692\\
19&   12.8728  &  -4.8958&50  &    0.1183  &   0.3532  &   0.5285\\
20&   14.5688&    -7.9380 &50 &    0.1182&     0.2682   &  0.6137\\
\hline
\end{tabular}
\end{center}
\end{table}

\subsection{Desired Quadcopter States}
\textbf{Dynamics of the Payload}: By applying Newton's second law the payload dynamics is given by
\begin{equation}
\label{PAYLOADSECONDLAW}
m_p
\begin{bmatrix}
       \ddot{x}_{p,i}\\
    \ddot{y}_{p,i}\\
     \ddot{z}_{p,i}\\
\end{bmatrix}
=-\dfrac{1}{m_p}\bigg[\sum_{i=1}^{N}\mathbf{F}_{Cord,i}+\mathbf{F}_{Aero,p}-m_pg\mathbf{K}\bigg]
\end{equation}
\nomenclature{$m_p$}{Payload mass}\nomenclature{$\mathbf{v}_p$}{Payload velocity}
where $N$ is the total number of cables (quadcopters), $m_p$ is the payload mass, $\mathbf{a}_p=\ddot{x}_{p,i}\mathbf{I}+\ddot{y}_{p,i}\mathbf{J}+\ddot{z}_{p,i}\mathbf{K}$ is the payload acceleration, and $\mathbf{F}_{Cord,i}$ is the force in cable $i$ connecting the payload to quadcopter $i$. It is assumed that cable $i$ acts as a spring, therefore,
\begin{equation}
\label{CABLESPRING}
\mathbf{F}_{Cord,i}=k_i(l_i(t)-l_{0,i})
\begin{bmatrix}
         \dfrac{x_{i}-x_p}{l_i}\\
         \dfrac{y_{i}-y_p}{l_i}\\
          \dfrac{z_{i}-z_p}{l_i}\\
\end{bmatrix}
\end{equation}
where $l_i=\sqrt{(x_i-x_p)^2+(y_i-y_p)^2+(z_i-z_p)^2}$ is the current length of cable $i$, $k_i$ is the stiffness of cable $i$, and $l_{0,i}=\sqrt{(x_i(0)-x_p(0))^2+(y_i(0)-y_p(0))^2+(z_i(0)-z_p(0))^2}$ is the free length of cable $i$. Notice that the force in cable $i$ ($F_{Cord,i}$) is in the direction of the line segment connecting the payload to quadcopter $i$.

Aerodynamic force exerted on the payload is given by
\begin{equation}
\mathbf{F}_{Aero,p}=
\begin{bmatrix}
         C_{p,x}\dot{x}_p\\
         C_{p,y}\dot{y}_p\\
         C_{p,z}\dot{z}_p\\
\end{bmatrix}
+\delta \mathbf{F}_{Aero,p}
\end{equation}
where $\delta \mathbf{F}_{Aero,p}$ is a zero-mean random vector with covariance matrix
\[
\mathbb{E}(\delta \mathbf{F}_{Aero,p}\delta \mathbf{F}_{Aero,p}^T)=
\begin{bmatrix}
       0.9985  &  0.0488    &0.0302\\
    0.0488 &   0.9906 &  -0.0390\\
    0.0302  & -0.0390    &0.9840  \\
\end{bmatrix}
.
\]

Because leaders' trajectories are linear with respect to time, follower quadcopter desired trajectories given by the homogeneous deformation are also linear with respect to time (see Eq. \eqref{followersdesired}.). To specify the payload acceleration used in Eq. \eqref{LQG Controller}, $x_i$, $y_i$, and $z_i$ in Eqs. \eqref{PAYLOADSECONDLAW} and \eqref{CABLESPRING} are substituted by
 \begin{equation}
 \begin{split}
 x_i(t)=x_{d,i}(t)+\delta x_i(t)\\
 y_i(t)=y_{d,i}(t)+\delta y_i(t)\\
 z_i(t)=z_{d,i}(t)+\delta z_i(t)\\
 \end{split}
\end{equation}
where $x_{d,i}(t)$, $y_{d,i}(t)$, and $z_{d,i}(t)$ are obtained from Eq. \eqref{followersdesired} and $\delta \mathbf{r}_i(t)=\delta x_i(t)\mathbf{I}+\delta y_i(t)\mathbf{J}+\delta z_i(t)\mathbf{K}$ is a zero-mean random vector with covariance
\[
\mathbb{E}(\delta \mathbf{r}_i\delta \mathbf{r}_i^T)=
\begin{bmatrix}
       0.9985  &  0.0488    &0.0302\\
    0.0488 &   0.9906 &  -0.0390\\
    0.0302  & -0.0390    &0.9840  \\
\end{bmatrix}
.
\]

\begin{figure}
\centering
\subfigure[]{\includegraphics[width=0.3\linewidth]{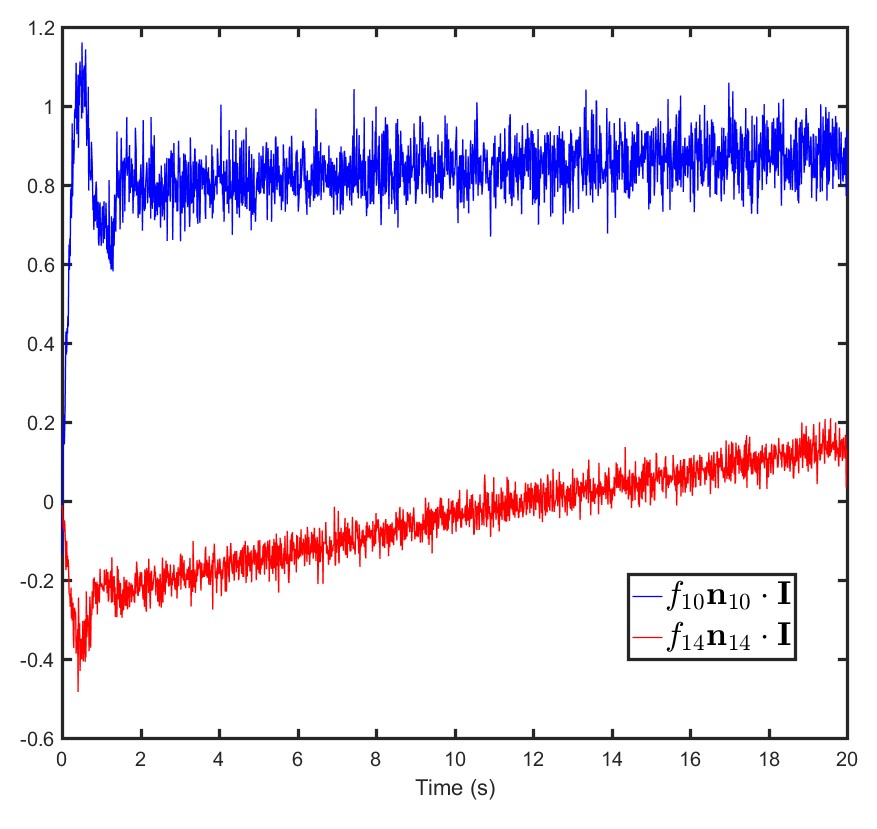}}
\subfigure[]{\includegraphics[width=0.3\linewidth]{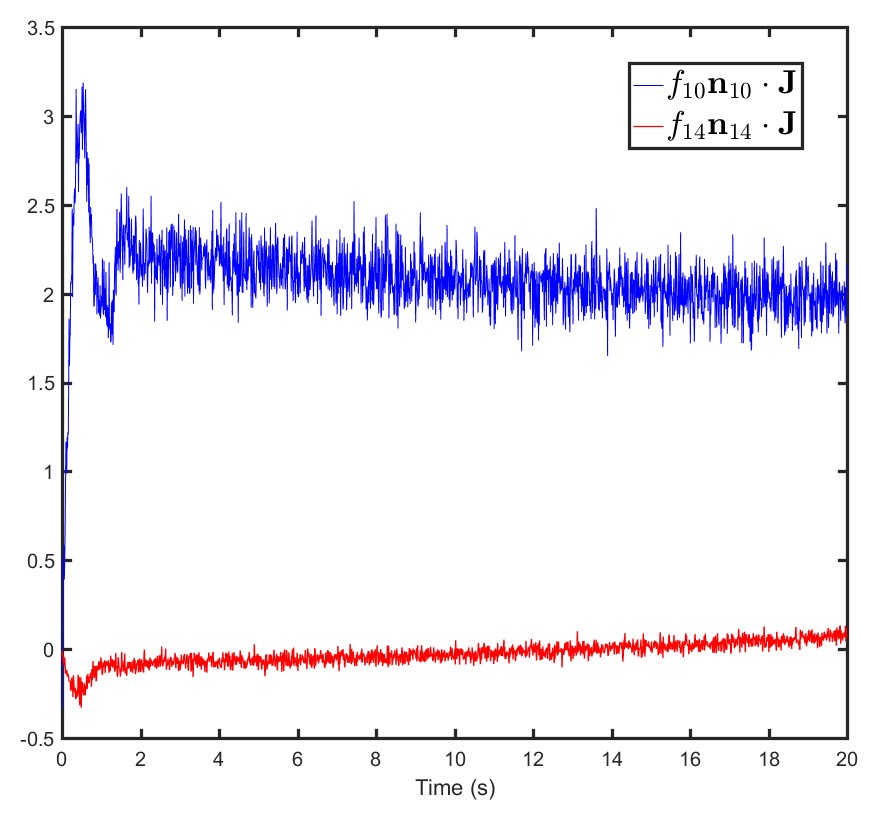}}
\subfigure[]{\includegraphics[width=0.3\linewidth]{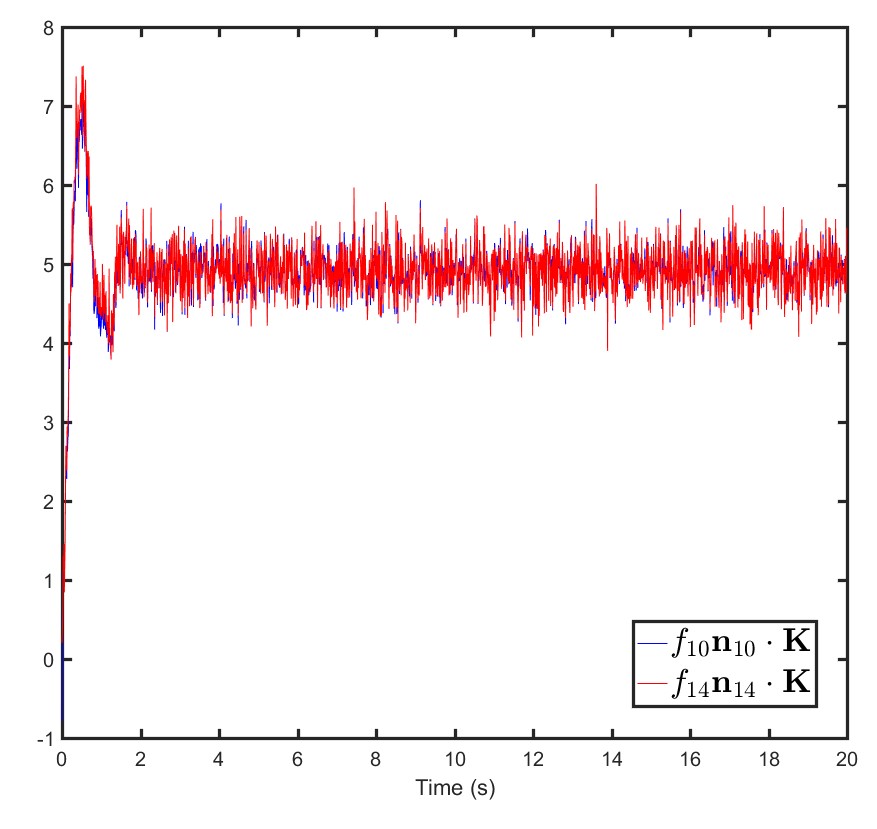}}
\caption{(a)$x$ components of the tensions generated in cables $10$ and $14$; (b)$y$ components of the tensions generated in cables $10$ and $14$; (c) $z$ components of the tensions generated in cables $10$ and $14$}
\label{f10Desired}
\end{figure}

 \textbf{Desired Tension Force $\mathbf{F}_{Cord,i}$}:  To assign matrices $A_{i,k}$ and $B_{i,k}$ (see Eq. \eqref{LQG Controller}), one must determine tension force $\mathbf{F}_{Cord,i}$ exerted on quadcopter $i$. Given the $i$th quadcopter position $\mathbf{r}_i$ and payload position $\mathbf{r}_p$,
 \begin{equation}
 \mathbf{n}_i=\dfrac{\mathbf{r}_i-\mathbf{r}_p}{\|\mathbf{r}_i-\mathbf{r}_p\|}
 \end{equation}
 assigns the direction of cable $i$. By applying Newton's second law, we can write
 \begin{equation}
 \label{PAYLOADDYNAMICS}
 \sum_{i=1}^{N}\mathbf{F}_{i}=\sum_{i=1}^{N}T_i\mathbf{n}_i=m_p(g\mathbf{K}+\ddot{\mathbf{r}}_p)
 \end{equation}
Note that $ \ddot{\mathbf{r}}_p=\ddot{x}_{p,i}\mathbf{I}+\ddot{y}_{p,i}\mathbf{J}+\ddot{z}_{p,i}\mathbf{K}$ is obtained from Eq. \eqref{PAYLOADSECONDLAW}, therefore, the right hand side of Eq. \eqref{PAYLOADDYNAMICS} as well as the unit vector 
\begin{equation}
\mathbf{n}_p=\dfrac{m_p(g\mathbf{K}+\ddot{\mathbf{r}}_p)}{\|m_p(g\mathbf{K}+\ddot{\mathbf{r}}_p)\|}
 \end{equation}
are known. By writing Eq.  \eqref{PAYLOADDYNAMICS}  component wise, we obtain three equations with twenty unknowns $f_1,\dots,f_{N}$. This is an over-determined problem that can be solved by modeling cables as parallel springs. We assume all cables have the same stiffness $k_i=k$, thus,
\begin{equation}
\label{ImportantEquality}
\dfrac{f_1}{d_1}=\dots=\dfrac{f_{N}}{d_{N}}
\end{equation}
where $\mathbf{d}_i=d_i\mathbf{n}_i$ is the axial displacement of the cable $i$. Payload is displaced in the direction of the vector $\mathbf{n}_p$.
Define $\mathbf{d}_p=d_p\mathbf{n}_p$ as the payload displacement,
\begin{equation}
d_i=\dfrac{d_p}{\mathbf{n}_i.\mathbf{n}_p}
\end{equation}
is substituted in Eq. \eqref{ImportantEquality}. Thus, Eq.  \eqref{ImportantEquality} can be rewritten as follows:
\begin{equation}
\label{ImportantEquality}
f_1\mathbf{n}_1.\mathbf{n}_p=\dots=f_{N}\mathbf{n}_{N}.\mathbf{n}_p.
\end{equation}
 Eq. \eqref{PAYLOADDYNAMICS} can be converted to the following with a dot product operation:
\begin{equation}
 \label{PAYLOADDYNAMICSs}
\sum_{i=1}^{N}f_{i}\mathbf{n}_i.\mathbf{n}_p=\|m_p(g\mathbf{K}+\ddot{\mathbf{r}}_p)\|.
 \end{equation}
Eqs. \eqref{ImportantEquality} and \eqref{PAYLOADDYNAMICSs} assigning  $f_1,~f_2, \dots,~f_{N}$ can then be written as follows:
\begin{equation}
\begin{bmatrix}
         \mathbf{n}_1\cdot\mathbf{n}_p&-\mathbf{n}_2\cdot\mathbf{n}_p&0&\dots&0&0\\
         0&\mathbf{n}_2\cdot\mathbf{n}_p&-\mathbf{n}_3\cdot\mathbf{n}_p&\dots&0&0\\
         \vdots&\vdots&\vdots&\ddots&\vdots&\vdots\\
         0&0&0&\dots&\mathbf{n}_{19}\cdot\mathbf{n}_p&-\mathbf{n}_{N}\cdot\mathbf{n}_p\\
         \mathbf{n}_1\cdot\mathbf{n}_p&\mathbf{n}_2\cdot\mathbf{n}_p&\mathbf{n}_3\cdot\mathbf{n}_p&\dots&\mathbf{n}_{19}\cdot\mathbf{n}_p&\mathbf{n}_{N}\cdot\mathbf{n}_p\\
\end{bmatrix}
\begin{bmatrix}
         f_1\\
         f_2\\
         \vdots\\
         f_{19}\\
         f_{N}\\
\end{bmatrix}
=
\begin{bmatrix}
         0\\
         0\\
         \vdots\\
        0\\
         \|m_p(g\mathbf{K}+\ddot{\mathbf{r}}_p)\|\\
\end{bmatrix}
\end{equation}
Payload motion simulation parameters are given in Table \ref{payloadpar}. In Fig. \ref{f10Desired} components of the tension force in cables $10$ and $14$ are shown versus time.

\begin{figure}
\centering
\subfigure[]{\includegraphics[width=0.3\linewidth]{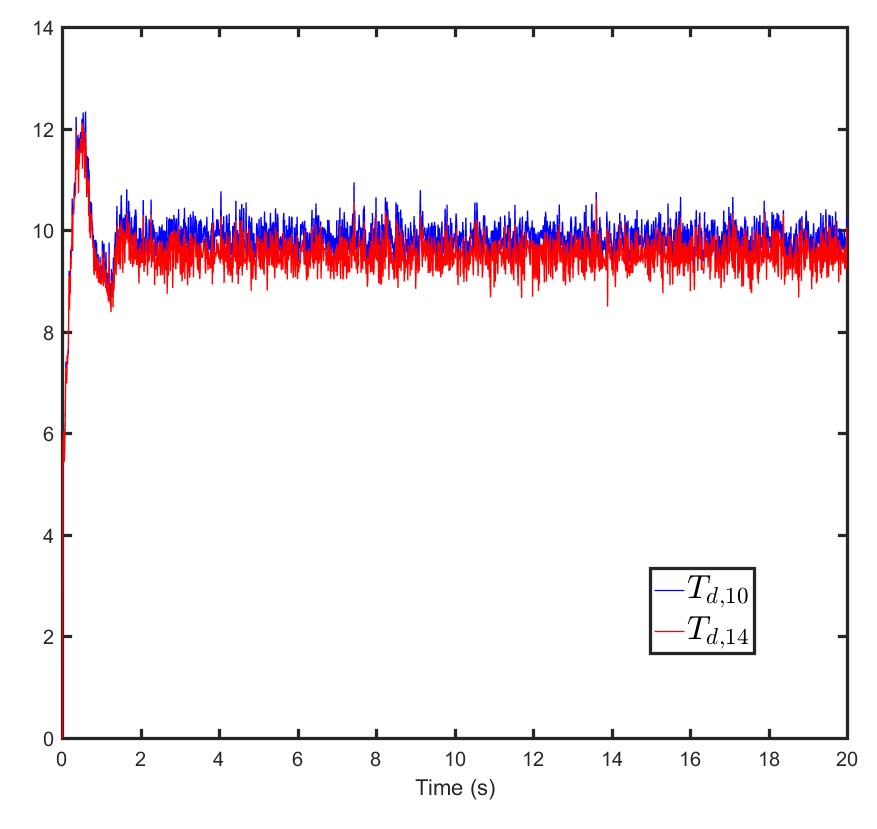}}
\subfigure[]{\includegraphics[width=0.3\linewidth]{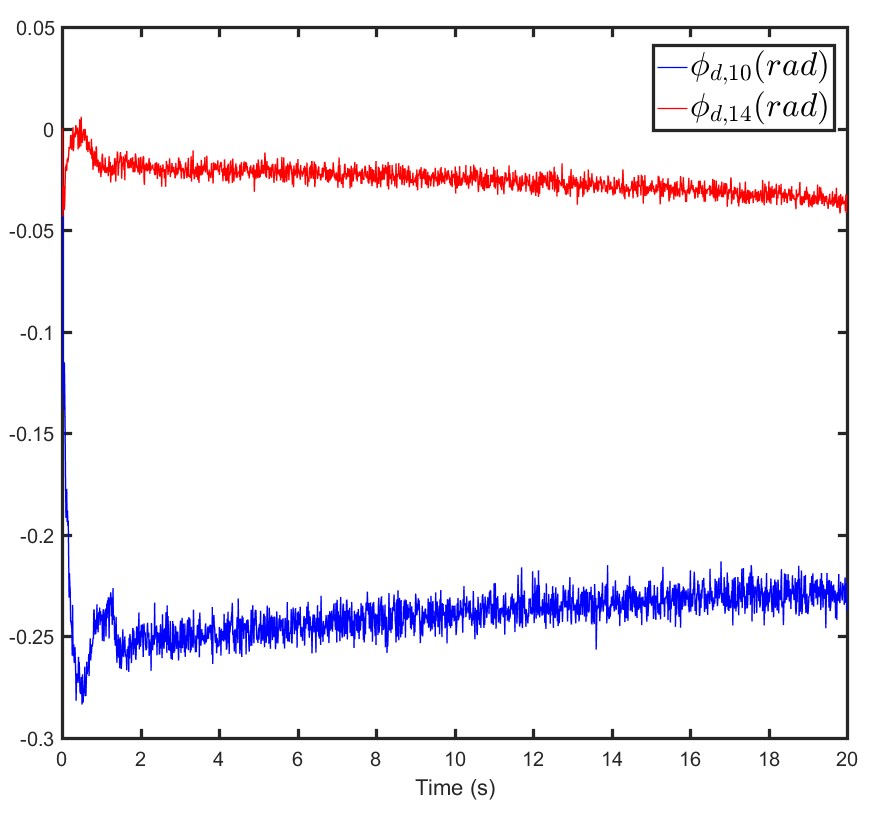}}
\subfigure[]{\includegraphics[width=0.3\linewidth]{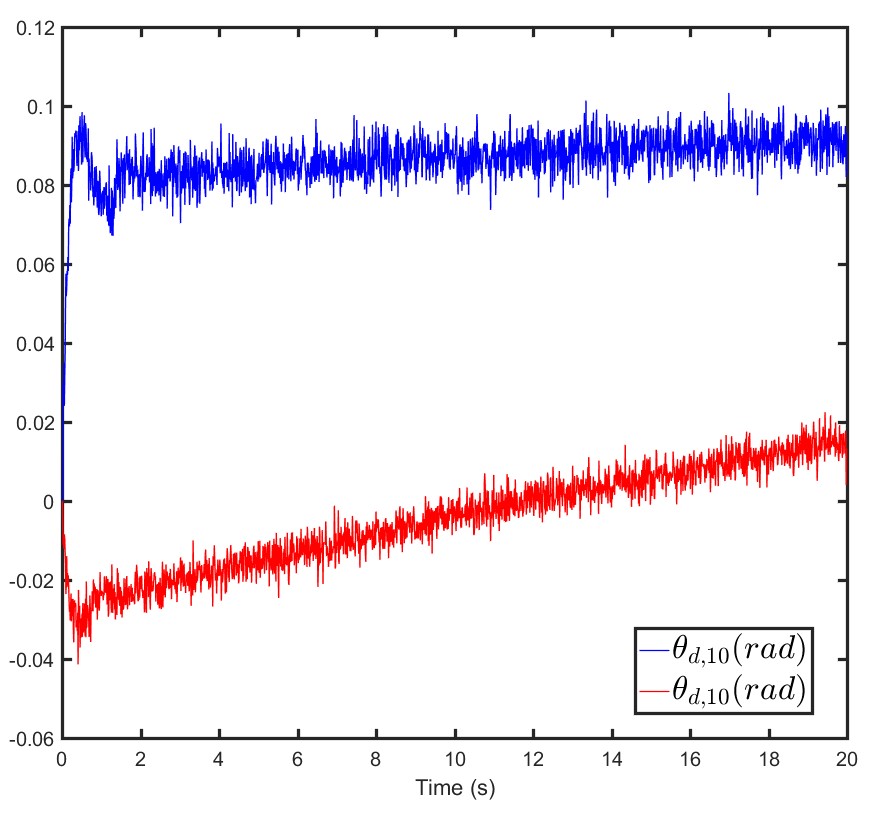}}
\caption{(a) Desired thrusts $T_{d,10}$ and  $T_{d,10}$ versus time; (b) Desired roll angles of quadcopters $10$ and $14$; (c) Desired pitch angles of quadcopters $10$ and $14$}
\label{phithetadesired}
\end{figure}

\textbf{Desired Thrust $T_{d,i}$}: Desired thrusts of the quadcopters are computed by using Eq. \eqref{THRUST}. Fig. \ref{phithetadesired}(a) depicts desired thrusts $T_{d,10}$ and $T_{d,14}$ versus time. 

 \textbf{Desired Euler Angles }: The desired path of quadcopter $i\in V$ is a straight line, so $\psi_{d,i}(t)=0$ ($t\in [0,2]$). Desired Euler angles of quadcopters $10$ and $14$ are calculated from Eq. \eqref{fidthetad} and shown versus time in Fig. \ref{phithetadesired}(b) and (c).
 \begin{table}[h]
     \centering
     \caption{Payload Motion Simulation Parameters }
     \begin{tabular}{c c}
     \hline
         Parameter & Value\\
         \hline
         \hline
         $m_p$&$10$ \\
         $k_i$&$100 kg/s^2$\\
         $N$&$20$ \\
         $C_{p,x}$&$4kg/s$\\
         $C_{p,y}$&$4kg/s$\\
         $C_{p,z}$&$4kg/s$\\
        \hline
     \end{tabular}
     \label{payloadpar}
 \end{table}
 
\subsection{LQG Controller}
Because desired velocities $u_{d,i}, v_{d,i}, w_{d,i}$ are constant, Euler angles $\phi_{d,i}, \theta_{d,i}, \psi_{d,i}$ and their derivatives  are small, and dynamics of  quadcopter $i$ can be approximated by linear time invariant (LTI) dynamics with $A_{i,k}=A_i$ and $A_{i,k}=B_i$. Given the desired quadcopter states at $t=10s$, matrices $A_1,\dots, A_6$ and $B_1,\dots,B_6$ are computed from Eq. \eqref{linearization}.

We consider $\Delta t=t_k-t_{k-1}=1$, therefore, the initial state $\delta{X}_{k,i,0}$ defined by Eq. \eqref{initialtimevarying} is updated at $t_{k-1}=1,\dots,19$ from \\

\[
\begin{split}
\forall k,\forall i\in V, H_{i,k}=E_{i,k}=0.01
\begin{bmatrix}
98 &    1 &   -3     &0 &    2&     0&     0     &0 &    3&    -2&    -7&     4\\
 1 &   91&    -1    & 0&    -4     &2 &   -1    & 3&    -4 &   -2    &-3 &    0\\
-3 &   -1    &98   & -4    &-2    & 0  &   0   & -5    &-3  &   5   & -1  &   2\\
 0 &    0   & -4  &  99   &  2   &  3   &  2  &   1   &  0   & -3  &  -5   &  2\\
 2 &   -4  &  -2 &    2  &  90  &   5    &-1 &    1  &   1    & 1 &    3    & 0\\
 0  &   2 &    0&     3 &    5 &   95     &1&     0 &    1     &0&    -1&    -6\\
 0  &  -1&     0     &2    &-1&     1&    97     &2&    -2&     0    &-4 &    4\\
 0  &   3    &-5    & 1    & 1    & 0 &    2    &97    & 5 &   -1    & 0  &   5\\
 3  &  -4   & -3   &  0   &  1   &  1  &  -2   &  5   & 98  &  -2   &  1   &  0\\
-2  &  -2  &   5  &  -3  &   1  &   0   &  0  &  -1  &  -2   & 96  &  -2    &-1\\
-7  &  -3 &   -1 &   -5 &    3 &   -1    &-4 &    0 &    1    &-2 &  101&    -1\\
 4  &   0&     2&     2&     0&    -6     &4&     5&     0    &-1&    -1 &   97\\
\end{bmatrix}
\end{split}
\]
to simulate quadcopter collective motion. 
 \begin{figure}
\centering
\subfigure[]{\includegraphics[width=0.45\linewidth]{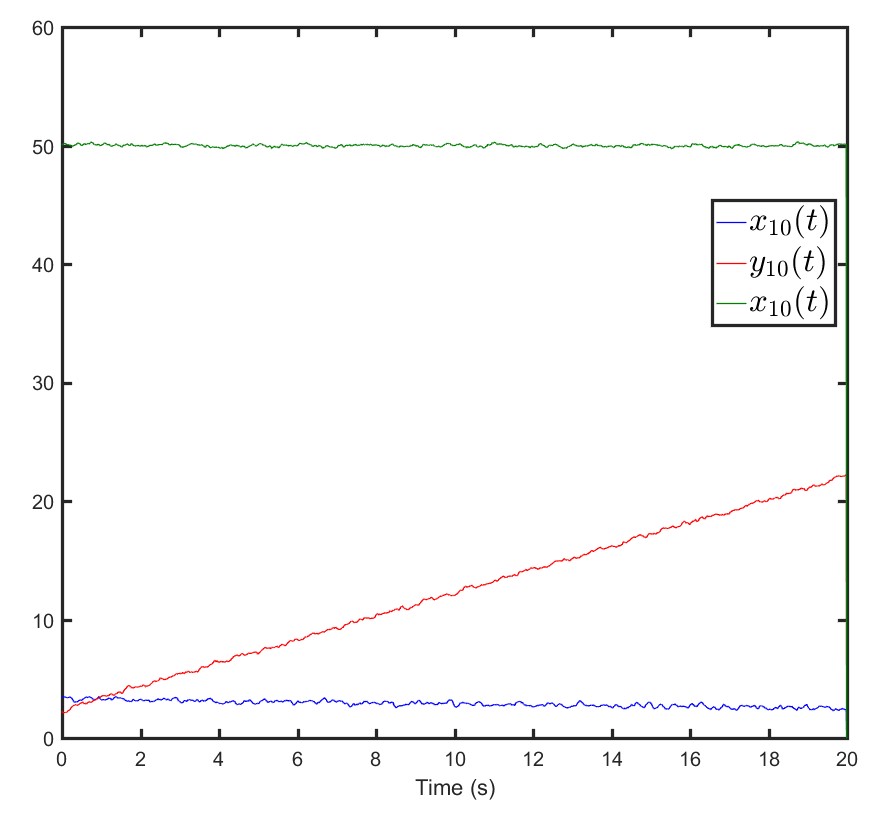}}
\subfigure[]{\includegraphics[width=0.45\linewidth]{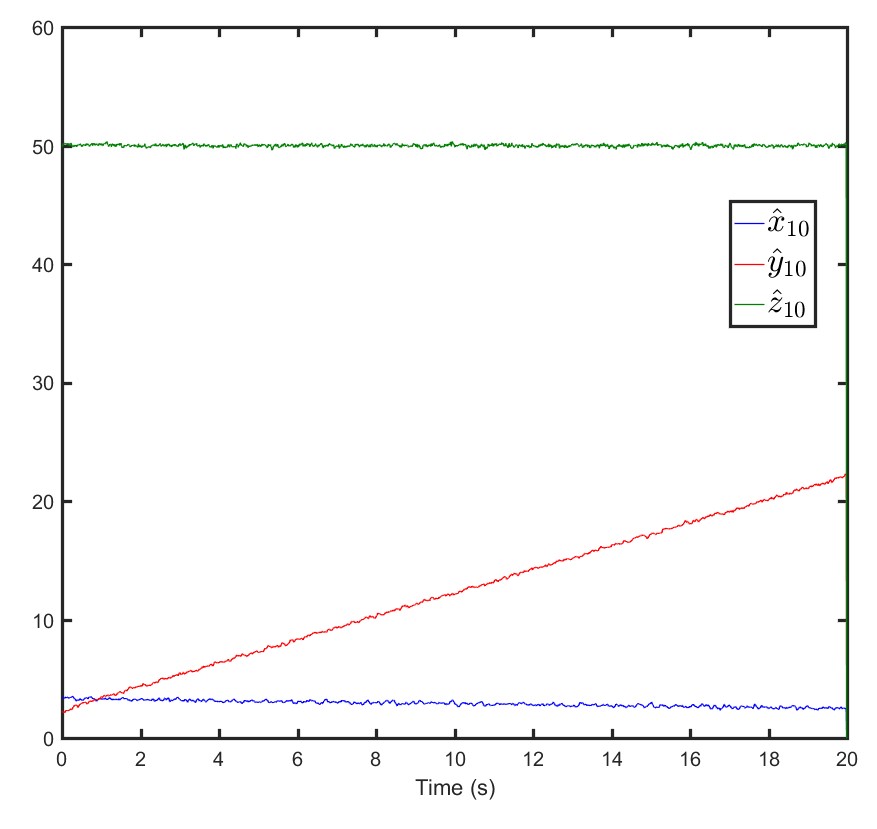}}
\subfigure[]{\includegraphics[width=0.45\linewidth]{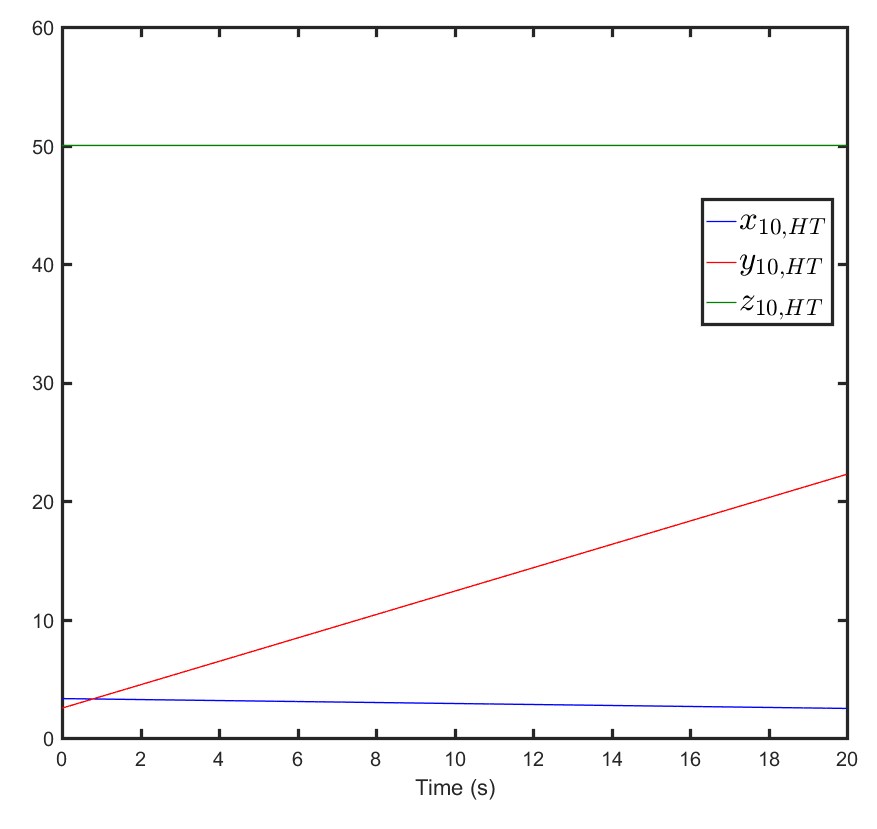}}
\caption{(a) Components of quadcopter $10$ actual position; (b) Components of quadcopter $10$ position estimated by the Kalman filter; (c) Components of quadcopter $10$  desired position as given by continuum deformation.}
\label{followerspositionsangles}
\end{figure}
\begin{figure}
\centering
\subfigure[]{\includegraphics[width=0.45\linewidth]{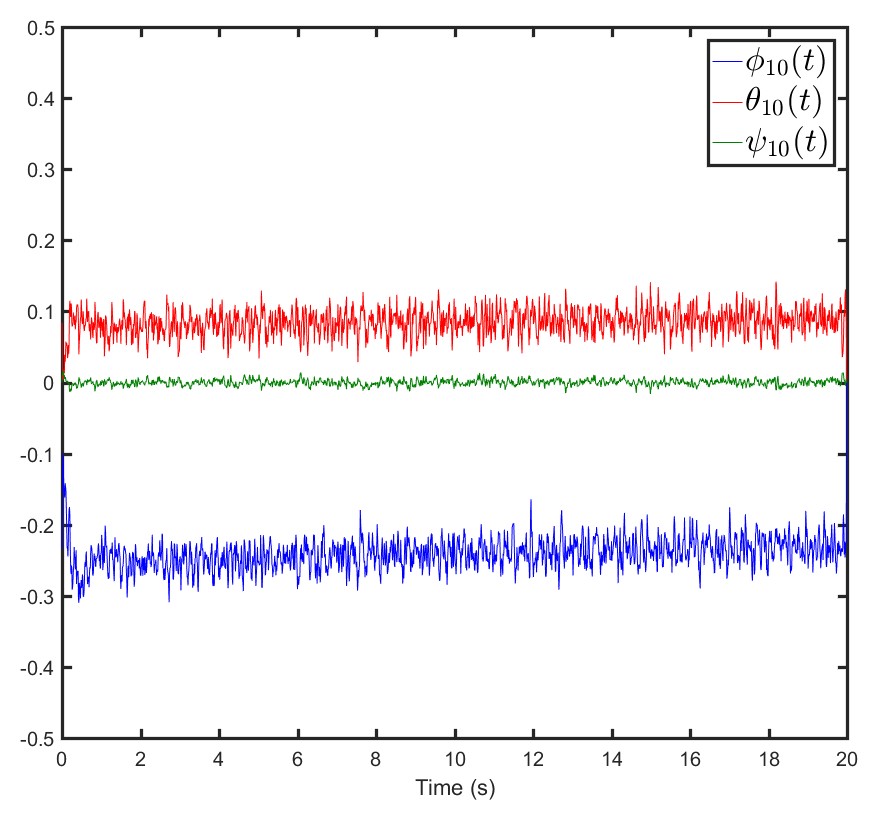}}
\subfigure[]{\includegraphics[width=0.45\linewidth]{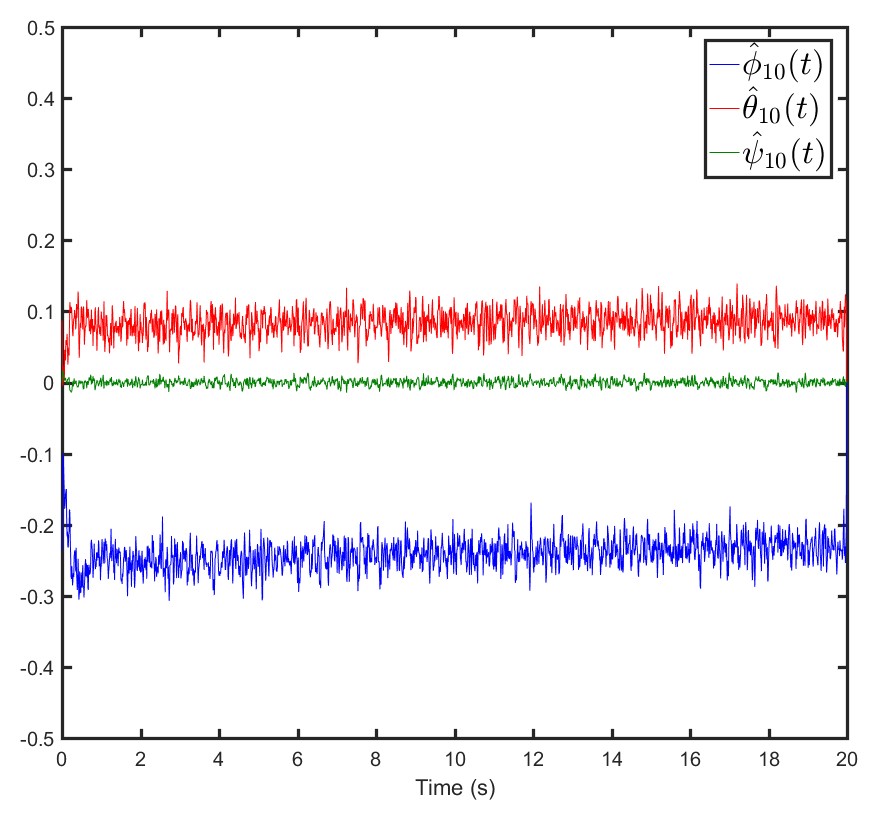}}
\caption{(a) Actual Euler angles of Quadcopter $10$; (b) Euler angles of Quadcopter $10$ estimated by the Kalman Filter.}
\label{EULERANGLES}
\end{figure}
\subsection{Results}
Fig. \ref{followerspositionsangles} shows $x_{10}$, $y_{10}$, and $z_{10}$ (components of actual position of the quadcopter  $10$), $\hat{x}_{10}$, $\hat{y}_{10}$, and $\hat{z}_{10}$ (Kalman filter estimation),  and $x_{d,10}$, $y_{d,10}$, and $z_{d,10}$ versus time.  Actual and estimated Euler angles of quadcopter $10$ are shown versus time in Fig. \ref{EULERANGLES}.
\begin{figure}
\center
\includegraphics[width=5.in]{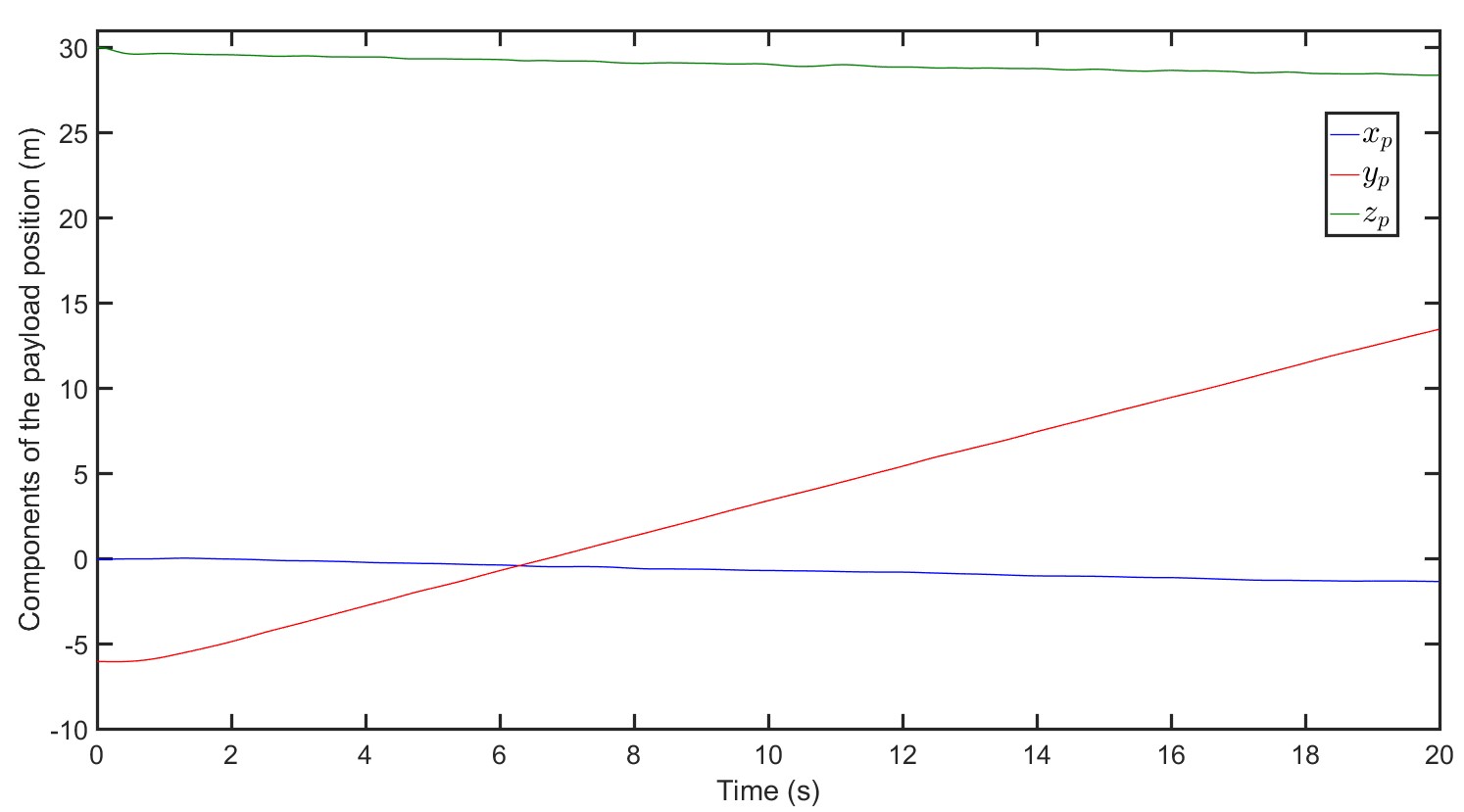}
\caption{Components of the payload position versus time.}
\label{paloadposition}
\end{figure}
\begin{figure}
\center
\includegraphics[width=5.in]{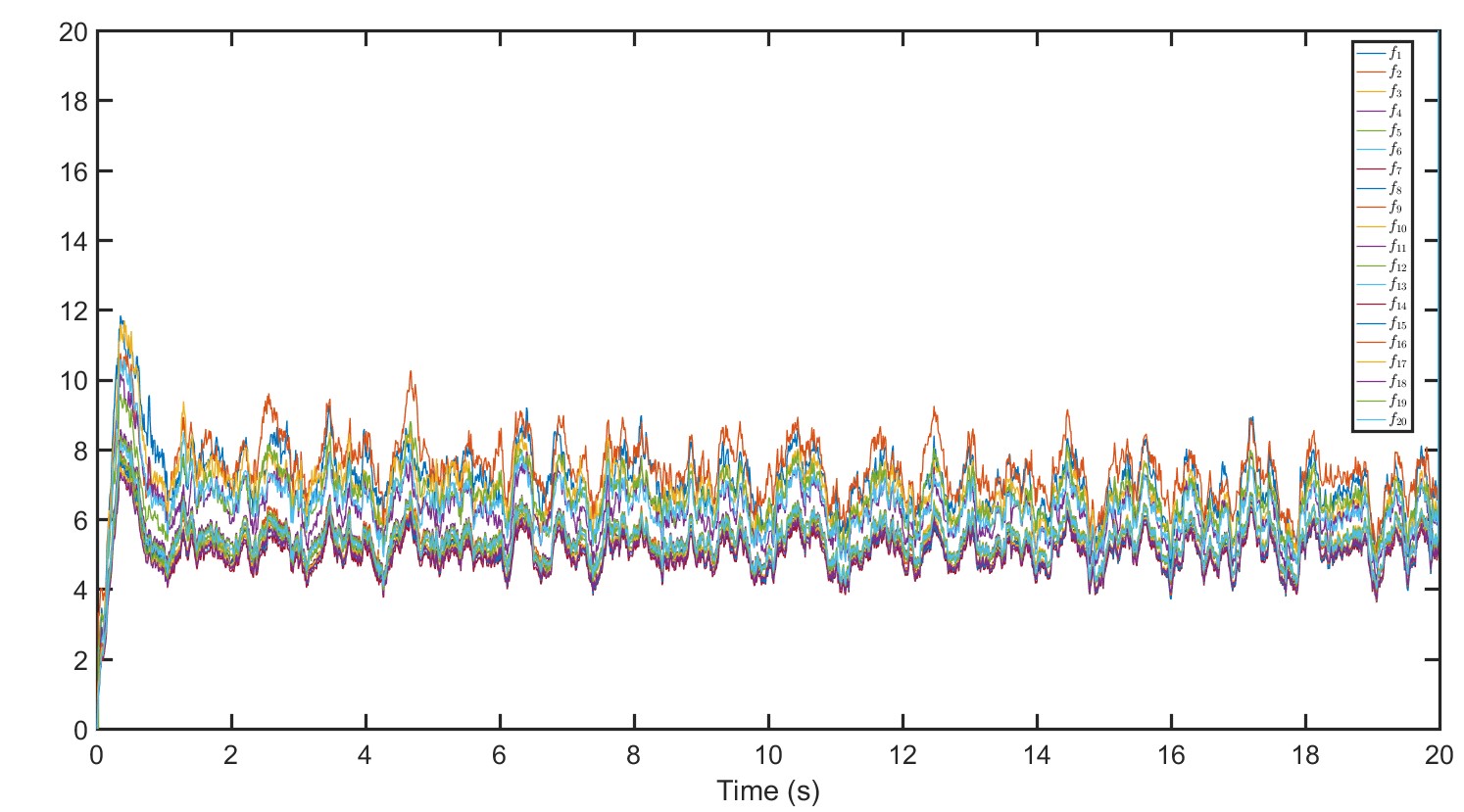}
\caption{Tension forces in the connecting cables.}
\label{tensioncables}
\end{figure}
Fig. \ref{paloadposition} shows payload position components ($x_p(t)$, $y_p(t)$, and $z_p(t)$) versus time. Tensions in the connecting cables are also plotted versus time in Fig. \ref{tensioncables}. 
The MQS formation and the payload at different sample times are shown in Fig. \ref{schematicpayloaddelivery}.
\begin{figure}[h]
\centering
\subfigure[$t=0s$]{\includegraphics[width=0.30\linewidth]{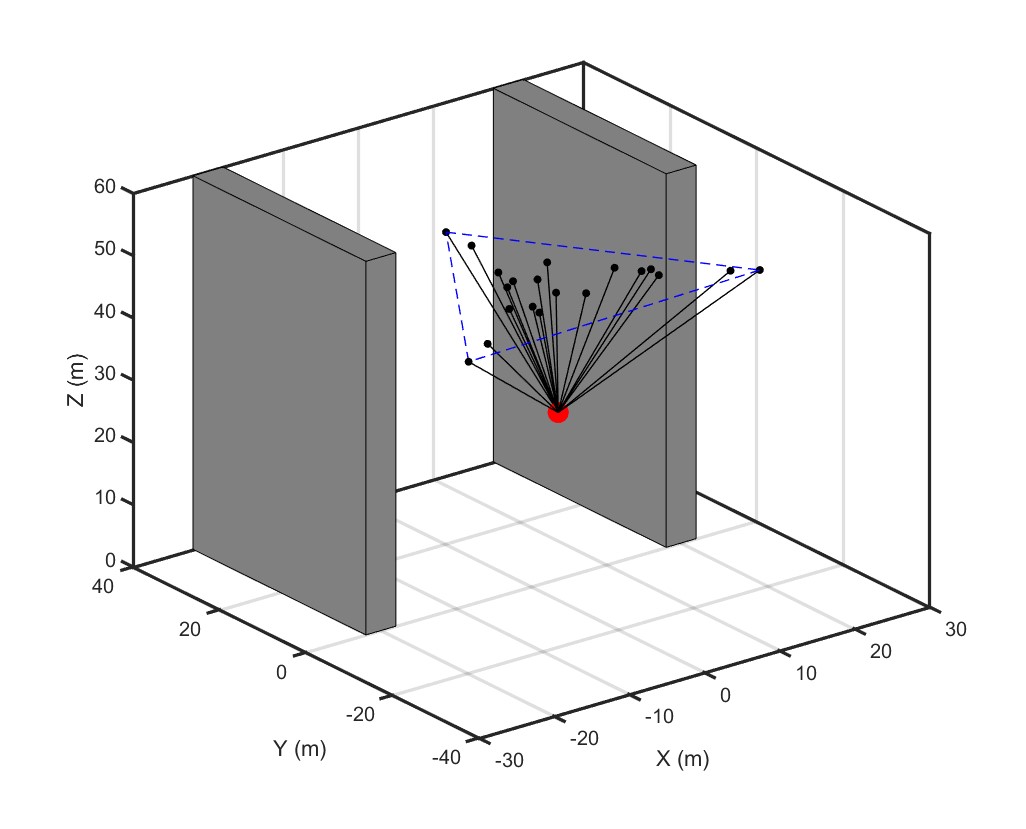}}
\subfigure[ $t=2.5s$]{\includegraphics[width=0.30\linewidth]{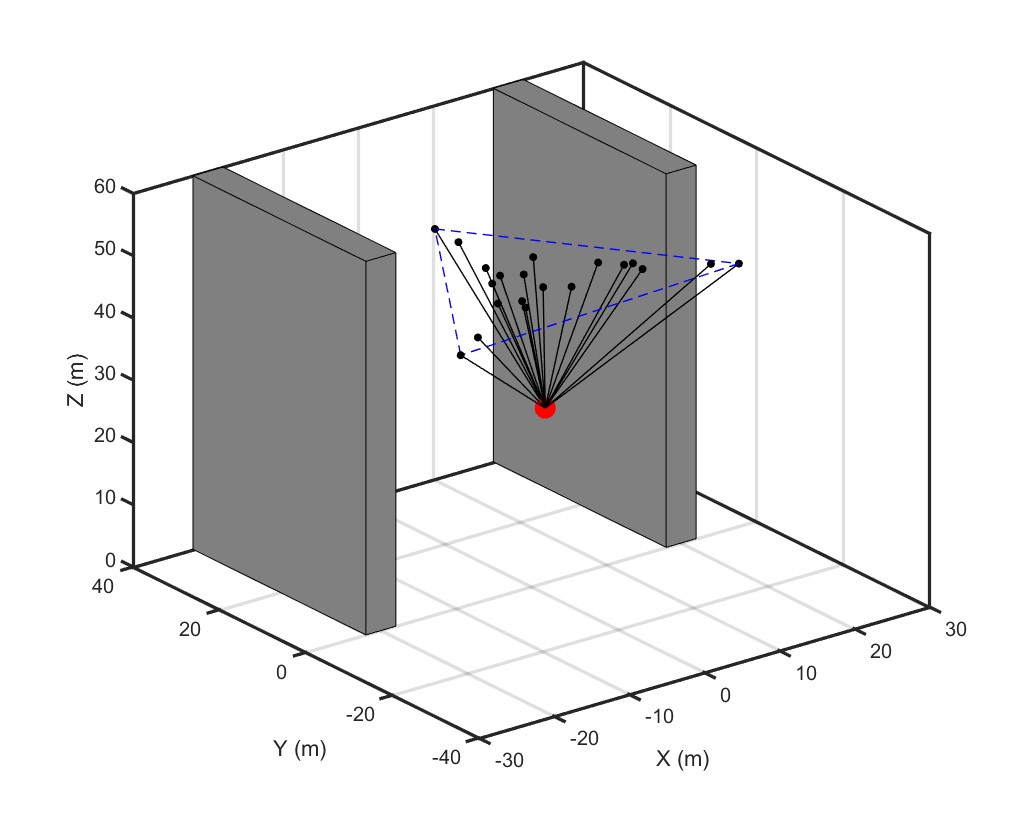}}
\subfigure[$t=5s$]{\includegraphics[width=0.30\linewidth]{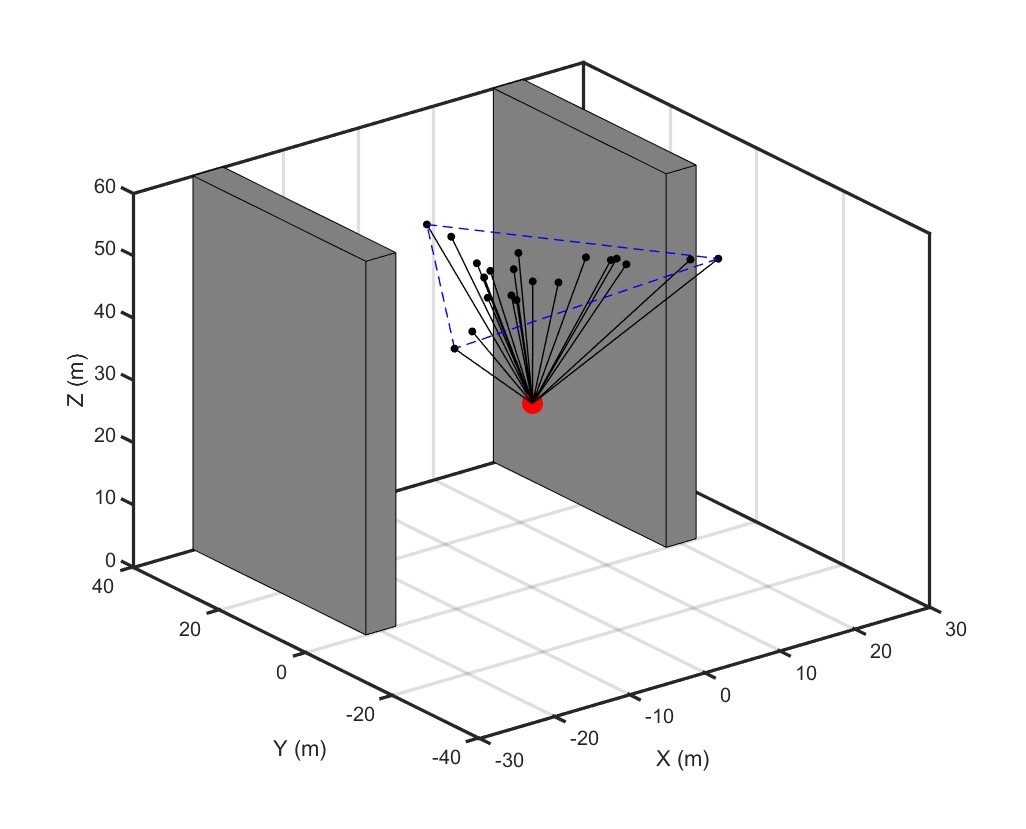}}
\subfigure[$t=7.5s$]{\includegraphics[width=0.30\linewidth]{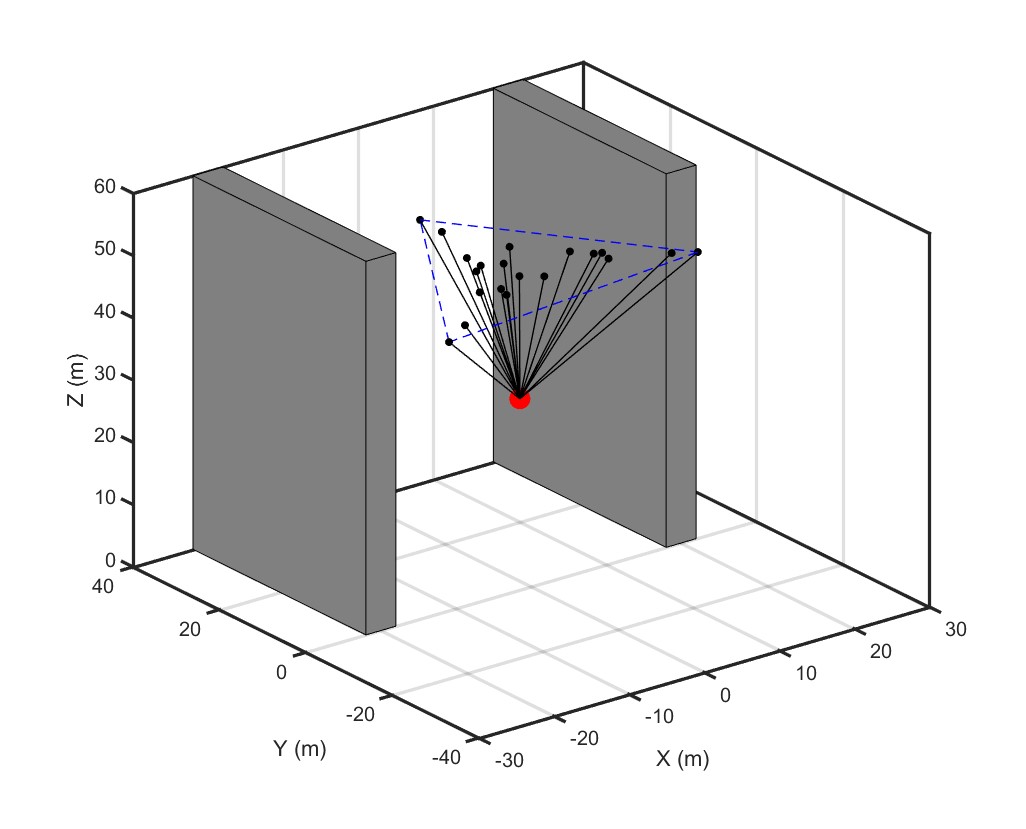}}
\subfigure[ $t=10s$]{\includegraphics[width=0.30\linewidth]{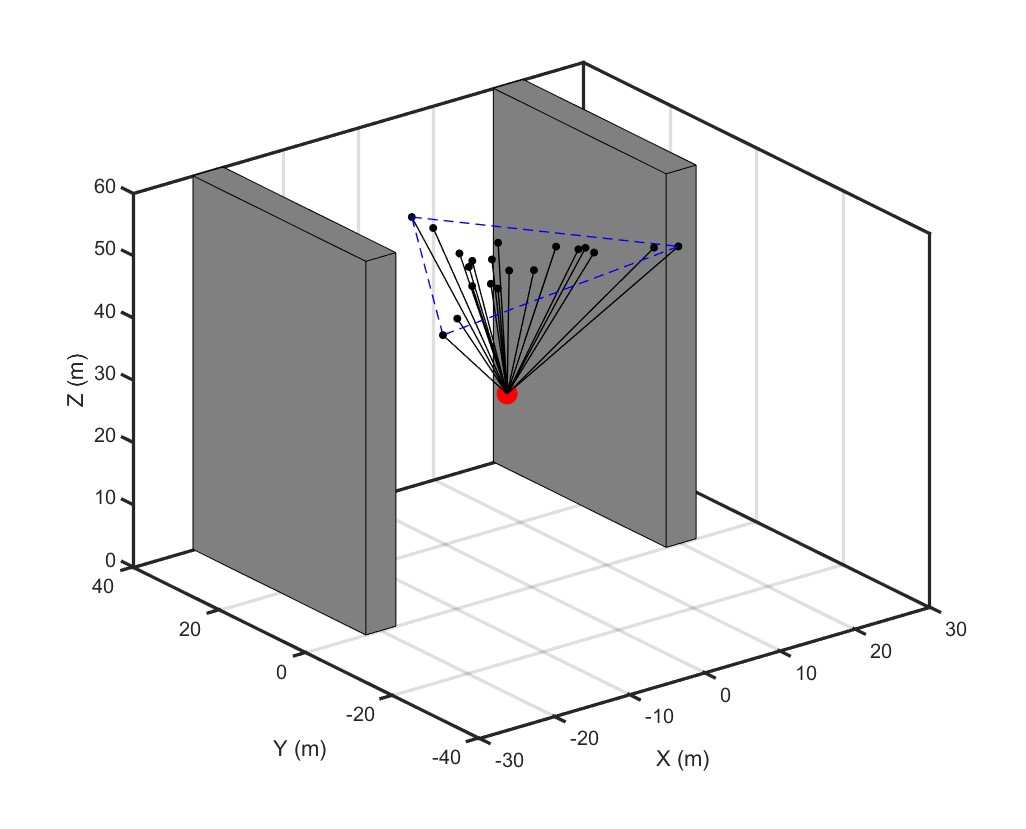}}
\subfigure[$t=12.5s$]{\includegraphics[width=0.30\linewidth]{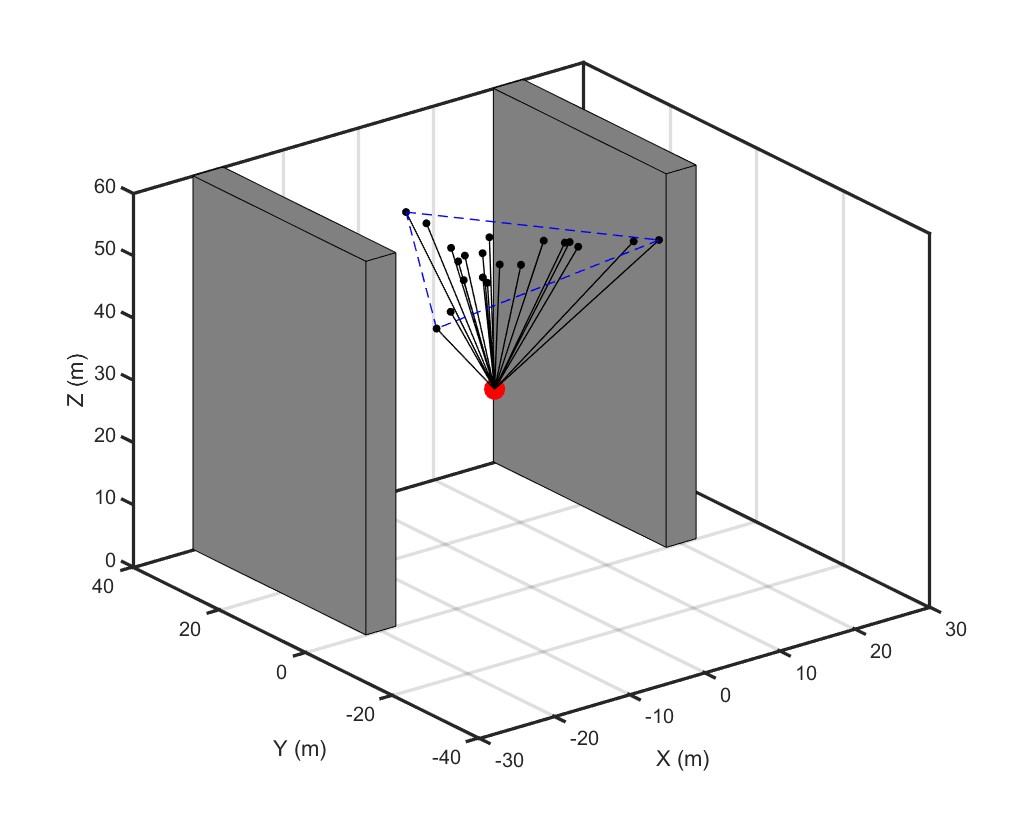}}
\subfigure[$t=15s$]{\includegraphics[width=0.30\linewidth]{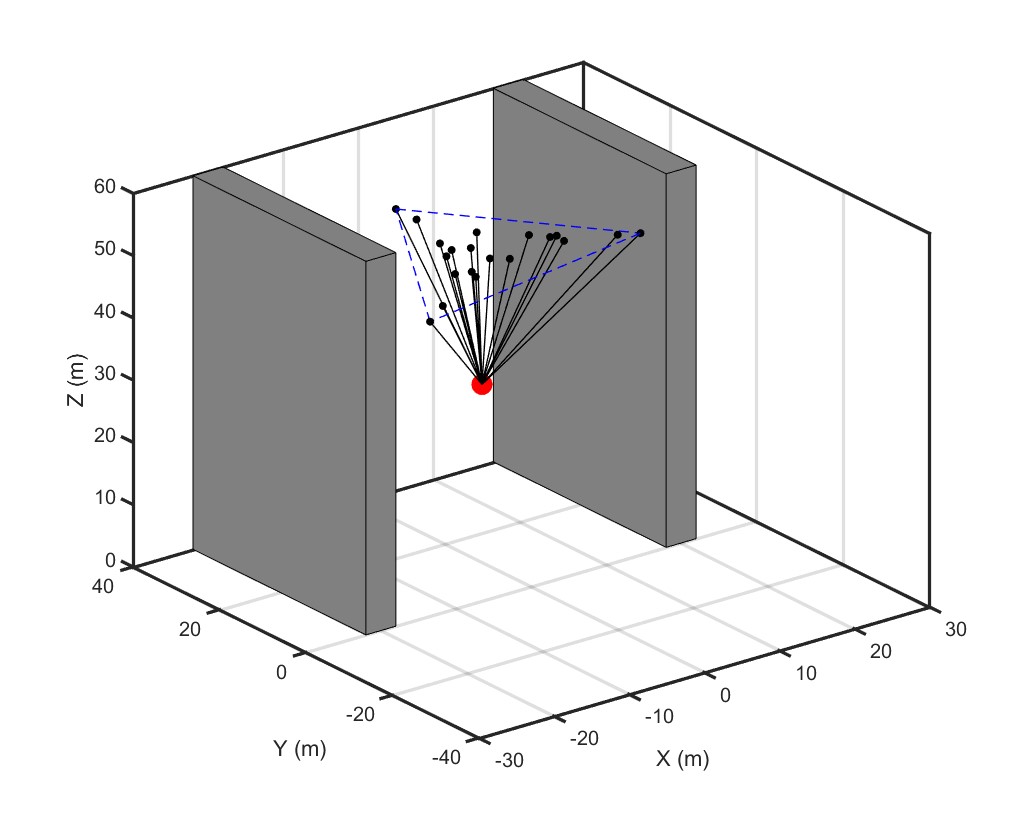}}
\subfigure[$t=17.5s$]{\includegraphics[width=0.30\linewidth]{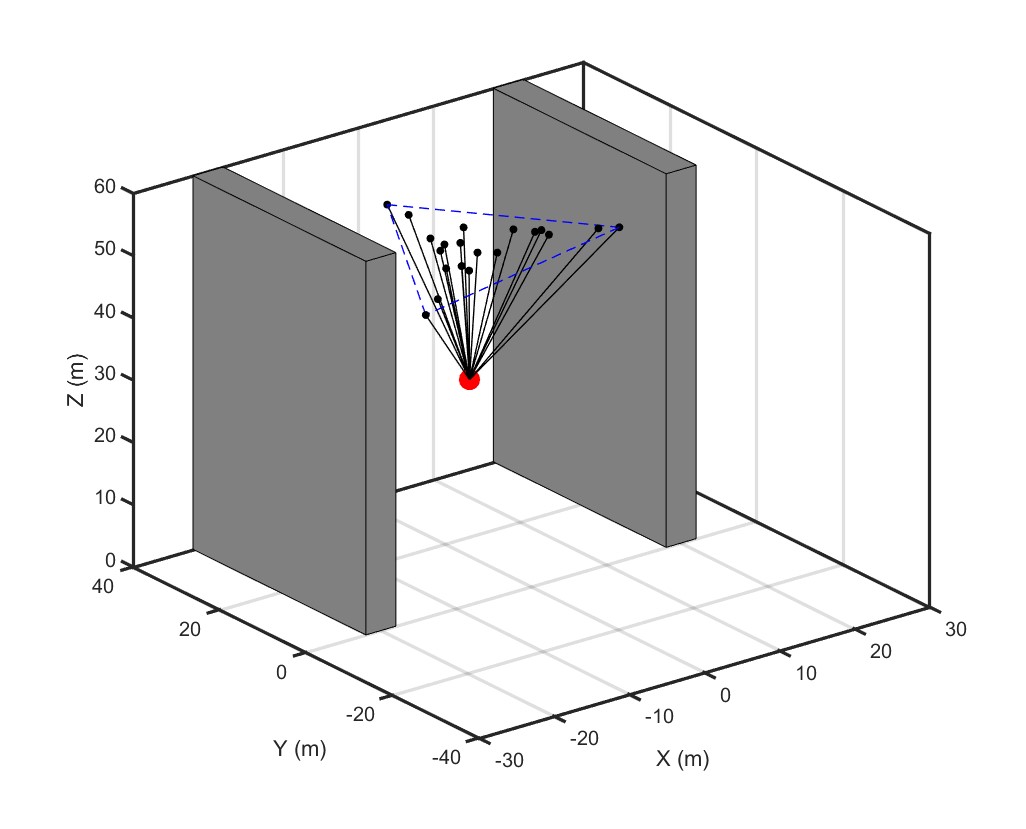}}
\subfigure[$t=20s$]{\includegraphics[width=0.30\linewidth]{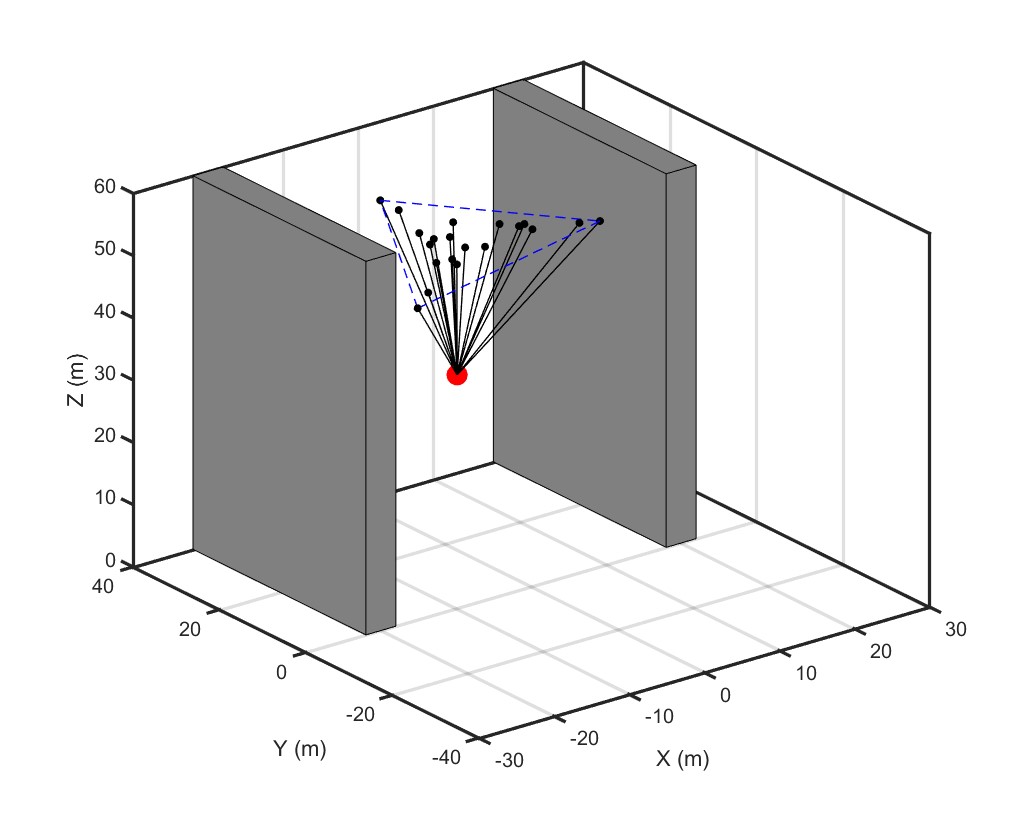}}
\caption{Payload and MQS configurations at Different Sample Times.}
\label{schematicpayloaddelivery}
\end{figure}

Fig. \ref{followerspositionsangles}(b) shows that the Kalman filter of quadcopter $10$ successfully estimates the actual position $\mathbf{r}_{10}(t)=x_{10}(t)\hat{\mathbf{I}}+y_{10}(t)\hat{\mathbf{J}}+z_{10}(t)\hat{\mathbf{K}}$ of quadcopter 10.  Fig.  \ref{followerspositionsangles}(c) shows that the controller is also able to accurately track the desired quadcopter $10$ trajectory. This situation is similar for all other quadcopters. 

The covariance matrix is defined as
\begin{equation}
Q_{\mathrm{Actual},10}=\sum_{i=1}^{2000}\dfrac{\bigg[X_{10}(0.01(i-1))-X_{d,10}(0.01(i-1))\bigg]\bigg[X(0.01(i-1))-X_d(0.01(i-1))\bigg]^T}{2000}.
\end{equation}
It is observed that eigenvalues of the symmetric matrix $Q_{\mathrm{Actual}}$ are all positive or zero. 
Additionally, deviation of the quadcopter $10$ from the desired position $\mathbf{r}_{10,HT}(t)$ is characterized by
\begin{equation}
Q_{\mathrm{pos},10}=\sum_{i=1}^{2000}\dfrac{
\begin{bmatrix}
x_{10}(0.01(i-1))-x_{d,10}(0.01(i-1))\\
y_{10}(0.01(i-1))-y_{d,10}(0.01(i-1))\\
z_{10}(0.01(i-1))-z_{d,10}(0.01(i-1))\\
\end{bmatrix}
\begin{bmatrix}
x_{10}(0.01(i-1))-x_{d,10}(0.01(i-1))\\
y_{10}(0.01(i-1))-y_{d,10}(0.01(i-1))\\
z_{10}(0.01(i-1))-z_{d,10}(0.01(i-1))
\end{bmatrix}
^T
}{2000}
\end{equation}
The simulation results show that
\begin{equation}
Q_{\mathrm{pos},10}=
\begin{bmatrix}
0.0173&   0.0292&    0.0627\\
0.0292&   0.2817&    0.5531\\
0.0627&   0.5531&    1.2605\\
\end{bmatrix}
.
\end{equation}
Expressing $Q_{\mathrm{pos},10}$ in its spectural decomposition form,
\begin{equation}
Q_{\mathrm{pos},10}=
\begin{bmatrix}
0.9978 &   0.0470   & 0.0462 \\
-0.0619 &   0.9099  &  0.4103\\
-0.0227  & -0.4122 &   0.9108\\
\end{bmatrix}
^T
\begin{bmatrix}
 0.0141 &        0 &        0\\
         0   & 0.0326&         0\\
         0    &     0 &   1.5128\\
\end{bmatrix}
\begin{bmatrix}
0.9978 &   0.0470   & 0.0462 \\
-0.0619 &   0.9099  &  0.4103\\
-0.0227  & -0.4122 &   0.9108\\
\end{bmatrix}
,
\end{equation}
it is concluded that the variance of the greatest deviation from the desired position is pointed to the direction $[0.0462~0.4103~09108]^T$. Note that $[0.0462~0.4103~09108]^T$ is along an axis that is almost parallel to the unit vector $\mathbf{K}$.  Also, variances of the greatest deviation in the $x-y$ deformation plane are not large.

Furthermore, deviation of actual Euler angles ($\phi_{10}(t),\theta_{10}(t),\psi_{10}(t)$) from the desired Euler angles ($\phi_{d,10}(t),\theta_{d,10}(t),\psi_{d,10}(t)$) is obtained from
\begin{equation}
Q_{\mathrm{Euler},10}=10^{-3}
\begin{bmatrix}
0.4538&    0.0168&    0.0153\\
0.0168&    0.2962&   -0.0432\\
0.0153&   -0.0432&    0.0173
\end{bmatrix}
\end{equation}
where 
\begin{equation}
Q_{\mathrm{Euler},10}=\sum_{i=1}^{2000}
\dfrac{
\begin{bmatrix}
\phi_{10}(0.01(i-1))-\phi_{d,10}(0.01(i-1))\\
\theta_{10}(0.01(i-1))-\theta_{d,10}(0.01(i-1))\\
\psi_{10}(0.01(i-1))-\psi_{d,10}(0.01(i-1))\\
\end{bmatrix}
\begin{bmatrix}
\phi_{10}(0.01(i-1))-\phi_{d,10}(0.01(i-1))\\
\theta_{10}(0.01(i-1))-\theta_{d,10}(0.01(i-1))\\
\psi_{10}(0.01(i-1))-\psi_{d,10}(0.01(i-1))
\end{bmatrix}
^T
}{2000}
.
\end{equation}
Notice that $\lambda_1(Q_{\mathrm{Euler},10})=0.0101\times 10^{-3}$, $\lambda_2(Q_{\mathrm{Euler},10})=0.3014\times 10^{-3}$, and $\lambda_3(Q_{\mathrm{Euler},10})=0.3014\times 10^{-3}$, thus, deviations of the actual Euler angles from their desired values are negligible. Considering the results of Eigen-analysis, it is concluded that the LQG controller can be successfully applied in a payload delivery mission when the payload is carried by large number of UAVs.
\section{Conclusion}
\label{Conclusion}
The paper applies the recently-developed continuum deformation algorithm to an application in which multiple quadcopters carry a single parcel cooperatively. Path planning and motion control algorithms are proposed and validated in simulation.  This paper described how a desired continuum deformation can be acquired by follower quadcopters only by knowing leaders' positions at certain sample times without any further communication, minimizing communication costs.  Because continuum deformation of an MQS is scalable, a large MQS team can cooperatively carry a heavy payload without the need for a heavy-lift quadcopter design. Because continuum deformation allows expansion and contraction of inter-agent distances, it also provides the ability for the team to pass through a narrow channel without collision as demonstrated in a twenty-quadcopter case study.  The quadcopter LQG controller provides robust tracking of the desired trajectory assigned by continuum deformation despite disturbance inputs. 

\section*{Acknowledgement}
This work was supported in part under Office of Naval Research (ONR) grant N000141410596.

\section*{References}
\bibliographystyle{aiaa}
\bibliography{referencee}

\end{document}